\documentstyle[aas2pp4]{article}

\begin{document}
\title{HELIUM ABUNDANCE IN THE MOST METAL-DEFICIENT 
BLUE COMPACT GALAXIES: I ZW 18 AND SBS 0335--052{\footnote[1]{The 
observations reported here were obtained at 
the Multiple Mirror Telescope Observatory, a joint facility of the Smithsonian
Institution and the University of Arizona, and at the W.M. Keck Observatory, 
which is operated as a scientific partnership among the California Institute of 
Technology, the University of California and the National Aeronautics and Space 
Administration.  The Observatory was made possible by the generous financial 
support of the W.M. Keck Foundation.}}}
\author{Yuri I. Izotov\footnote[2]{Visiting astronomer, 
National Optical Astronomical Observatories.}}
\affil{Main Astronomical Observatory, Ukrainian National Academy of Sciences,
Goloseevo, Kiev 252650, Ukraine \\ Electronic mail: izotov@mao.kiev.ua}
%\and
\author{Frederic H. Chaffee}
\affil{W. M. Keck Observatory, 65-1120 Mamalahoa Hwy., Kamuela,
HI 96743 \\ Electronic mail: fchaffee@keck.hawaii.edu}
%\and
\author{Craig B. Foltz}
\affil{Multiple Mirror Telescope Observatory, University of Arizona, 
Tucson, AZ 85721 \\ Electronic mail: cfoltz@as.arizona.edu}
%\and
\author{Richard F. Green}
\affil{National Optical Astronomy Observatories, 
Tucson, AZ 85726 \\ Electronic mail: rgreen@noao.edu}
%\and
\author{Natalia G. Guseva}
\affil{Main Astronomical Observatory, Ukrainian National Academy of Sciences,
Goloseevo, Kiev 252650, Ukraine \\ Electronic mail: guseva@mao.kiev.ua}
\and
\author{Trinh X. Thuan}
\affil{Astronomy Department, University of Virginia, Charlottesville, VA 22903
\\ Electronic mail: txt@virginia.edu}

\begin{abstract}

We present high-quality spectroscopic observations of the two most-metal 
deficient blue compact galaxies known, I Zw 18 and SBS 0335--052. We use the 
data to determine the heavy element and helium abundances. 
The oxygen abundances in the NW and the SE components of I Zw 18 are found to be
the same within the errors, 7.17$\pm$0.03 and 7.18$\pm$0.03 respectively,
although marginally statistically significant spatial variations of oxygen abundance
might be present. In contrast, we find a statistically significant gradient of
oxygen abundance in SBS 0335--052. The largest oxygen abundance,
12 + log O/H = 7.338$\pm$0.012, is found in the region 0\farcs6 to the
NE of the brightest part of the galaxy, and it decreases toward the
SW to values of $\sim$ 7.2, comparable to that in I Zw 18. 
The underlying stellar absorption
strongly influences the observed intensities of He I emission lines in the 
brightest NW component of I Zw 18, and hence this component 
{\it should not be used} for primordial He abundance determination. 
The effect of underlying
stellar absorption, though present, is much smaller in the SE component.
Assuming all systematic uncertainties are negligible, 
the He mass fraction $Y$ = 0.243$\pm$0.007 derived in this component is
in excellent agreement with recent measurements by Izotov \& Thuan, suggesting
the robustness of the technique applied in measurements of the helium abundance
in low-metallicity blue compact galaxies.

The high signal-to-noise ratio spectrum ($\geq$ 100 in the continuum)
of SBS 0335--052 allows us to measure the helium mass fraction with a precision
better than 2\% -- 5\% in nine different regions along the slit. 
We show that, while underlying
stellar absorption in SBS 0335--052 is important only for the He I $\lambda$4471
emission line, other mechanisms such as collisional and fluorescent enhancements 
are influencing the intensities of all He I emission
lines and should be properly taken 
into account. When the electron number density derived from [S II]
emission lines is used in SBS 0335--052, the correction of He I emission lines 
for collisional enhancement leads to systematically different He mass fractions 
for different He I emission lines. This {\it unphysical} result implies that
the use of the electron number density derived from [S II] emission lines,
being characteristic of the S$^+$ zone, but not of the
He$^+$ zone, will lead to an {\it incorrect} inferred value of $Y$. In the case 
of SBS 0335--052 it leads to a significant 
underestimate of the He mass fraction. In contrast, the self-consistent method
using the five strongest He I emission lines in the 
optical spectrum for correction
for collisional and fluorescent enhancements shows excellent agreement of the He
mass fraction  derived from the He I $\lambda$5876 and He I $\lambda$6678 
emission lines in all 9 regions of SBS 0335--052 used for the He abundance
determination. 

Assuming all systematic uncertainties are negligible, 
the weighted mean He mass fraction in SBS 0335--052 is
$Y$ = 0.2437$\pm$0.0014 when the three He I $\lambda$4471, $\lambda$5876 and 
$\lambda$6678 emission lines are used, and is 0.2463$\pm$0.0015 when the He I
$\lambda$4471 emission line is excluded. These values are in very good 
agreement with recent measurements of the He mass fraction in SBS 0335--052
by Izotov and coworkers. The weighted mean helium mass fraction in the two most 
metal-deficient BCGs I Zw 18 and SBS 0335--052, $Y$=0.2462$\pm$0.0015,
after correction for the stellar He production results in a primordial
He mass fraction $Y_p$ = 0.2452$\pm$0.0015. The derived $Y_p$
leads to a baryon-to-photon ratio of 
(4.7$^{+1.0}_{-0.8}$)$\times$10$^{-10}$ and to a
baryon mass fraction in the Universe $\Omega_b$$h^2_{50}$ = 
0.068$^{+0.015}_{-0.012}$,
consistent with the values derived from the primordial D and $^7$Li abundances, 
and supporting the standard big bang 
nucleosynthesis theory. For the most consistent set of primordial D, $^4$He,
and $^7$Li abundances we derive an equivalent number of light neutrino
species $N_\nu$ = 3.0$\pm$0.3 (2$\sigma$).

\end{abstract}


\section {INTRODUCTION}

   Blue compact galaxies (BCGs) are ideal objects for the determination
of the primordial helium abundance and hence 
of one of the fundamental cosmological parameters -- the baryon mass fraction
in the Universe. In the standard big bang model of nucleosynthesis (SBBN),
light isotopes D, $^3$He, $^4$He and $^7$Li were produced by nuclear reactions
a few seconds after the birth of the Universe. Given the number of
light neutrino species $N_\nu$ = 3 and the neutron lifetime, 
the abundances of
these light elements depend on one cosmological parameter only, the
baryon-to-photon ratio $\eta$, which in turn is directly related to the
density of ordinary baryonic matter $\Omega_b$
(Walker et al. 1991; Copi, Schramm \& Turner 1995; Sarkar 1996). 
The precise value of the 
primordial $^4$He abundance can be used to put a stringent limit on
the number of light neutrino species $N_\nu$ (Steigman, Schramm, \& Gunn 1977;
Burles et al. 1999; Lisi, Sarkar \& Villante 1999).

     The primordial helium mass fraction $Y_p$ of $^4$He is usually derived
by extrapolating the $Y$ -- O/H and $Y$ -- N/H correlations to O/H =
N/H = 0, as proposed originally by Peimbert \& Torres-Peimbert (1974, 1976)
and Pagel, Terlevich \& Melnick (1986). Many attempts at determining
$Y_p$ have been made, using these correlations on
various samples of dwarf irregular and BCGs (see
references in Pagel et al. (1992), Izotov, Thuan \& Lipovetsky (1994,
1997, hereafter ITL94 and ITL97)) and Izotov \& Thuan (1998ab). 
These dwarf systems are the least chemically
evolved galaxies known, so they contain very little helium manufactured by
stars after the big bang. Because the big bang production
of $^4$He is relatively insensitive to the density of matter, the primordial
abundance of $^4$He needs to be determined  to a very high precision 
(better than few percent) in order to put useful constraints on $\Omega_b$. 
This precision
can be achieved with very high signal-to-noise ratio 
optical spectra of BCGs. These
BCGs are undergoing intense bursts of star formation, giving birth to high
excitation supergiant H II regions and allowing an accurate determination of 
the
helium abundance in the ionized gas through the BCG's emission-line spectrum.

Recent conflicting $Y_p$ determinations have been made by two main groups using 
large samples of BCGs. One group (ITL94, ITL97, Izotov \& Thuan 1998a) finds
$Y_p$ $\sim$ 0.245 while the other (Olive \& Steigman 1995; Olive, 
Skillman \& Steigman 1997, hereafter OSS97) finds consistently lower 
values, with $Y_p$ in the 0.230 -- 0.234 range. There are several possible 
sources of systematic effects which could account for this disagreement.
Some of these arise from the differing quality of data
and different methods used for the He abundance determination.
ITL94, ITL97 and Izotov \& Thuan (1998a) used new high signal-to-noise 
spectroscopic observations of BCGs reduced in a homogeneous way.
They have applied a self-consistent method to correct He I emission
line intensities for several effects (collisional and fluorescent
enhancement, underlying stellar absorption)
which lead to deviation from values predicted
by recombination theory. The main assumptions of this technique are the
correctness of input atomic data for He I transitions and the requirement
that, in low-density supergiant H II regions, the intensity ratios of all 
He I emission lines after correction
must be best fitted by recombination values. The use of several He I
lines allows one to discriminate between collisional and fluorescent
enhancements which change line intensities in different ways. It also
allows one to estimate the importance of underlying stellar absorption
in each galaxy and to improve the precision of $Y_p$
determination due to the use of several lines. The details of this
approach are discussed by ITL97 and Izotov \& Thuan (1998a).

On the other hand, Olive \& Steigman (1995) and OSS97 use 
a more heterogeneous sample of BCGs observed by
different observers on different telescopes. Some of these data were
obtained many years ago with non-linear detectors. Therefore,
these galaxies should be
reobserved, especially those with the lowest metallicities. The approach
of OSS97 is based mainly on He mass fractions derived from the 
single He I $\lambda$6678 emission line and 
correction is made only for collisional enhancement using the
electron number density derived from the [S II]$\lambda$6717/$\lambda$6731
emission line ratio. This approach has several shortcomings. First,
one cannot discriminate between different mechanisms of He I emission lines
enhancement when a single He I emission line is used. 
For instance, fluorescent enhancement is neglected, while
it might be present. Second, at low electron number densities the determination 
of $N_e$ from [S II] emission lines is very
uncertain and in the majority of cases $N_e$ is arbitrarily set equal to 
the upper limit of 100 cm$^{-3}$. This assumption leads to lower $^4$He 
abundance in the SE component of I Zw 18 (Izotov \& Thuan 1998b) because the
true electron number density in this H II region is lower. 
More importantly, the use of $N_e$
derived from [S II] emission lines is {\it physically unreasonable}
because the S$^+$ and He$^+$ regions are not 
expected to be coincident due to the large 
difference in S and He ionization potentials. In some cases
this assumption leads to an {\it incorrect} result. For instance, in the dense
and hot H II region of SBS 0335--052 the use of $N_e$(S II) results
in a systematic underestimate of the helium mass fraction by 5 -- 10\%.

Because of its nearly primordial nature, many determinations
of the helium abundance have been carried out for I Zw 18. Most recently, Pagel 
et al. (1992), Skillman \& Kennicutt (1993, hereafter SK93) 
and ITL97 have derived a very low helium mass fraction for the NW component 
of I Zw 18: $Y$ = 0.228--0.231.
    ITL97 have attributed this abnormally low $Y$ to underlying stellar 
absorption
affecting all He I lines in I Zw 18. They cite as evidence the implausibly small
$Y$ $\leq$ 0.200 derived from the He I $\lambda$4471 line in I Zw 18, which is 
comparatively more affected by stellar absorption than the He I $\lambda$6678
line
because of its relatively lower equivalent width. Although the He I 
$\lambda$5876 line in the NW component of I Zw 18 is stronger than the He I 
$\lambda$6678 line, its intensity is affected by Galactic
interstellar sodium absorption and therefore cannot be used for $^4$He 
abundance determination. The value of $Y$ derived from the He I $\lambda$6678
emission line without correction for underlying stellar absorption is thus 
underestimated by at least 5\%. In contrast, Izotov \& Thuan (1998b) have 
derived a helium mass fraction $Y$ in the SE component of I Zw 18 of 
0.242$\pm$0.009, higher than the value $Y$ = 0.230$\pm$0.009 derived by SK93 
and $Y$ = 0.231$\pm$0.010 by Pagel et al. (1992) for the same component. 
They argued that this value should be adopted for $Y_p$. Recently
V\'ilchez \& Iglesias-P\'aramo (1998) have derived a mean value $Y$ = 
0.242$\pm$0.010 in I Zw 18 when the central knots are excluded,
confirming the high $Y$ value.
As for SBS 0335--052, Izotov et al. (1997a) and Izotov \& Thuan (1998a)
have found $Y$ = 0.245$\pm$0.006 and 0.249$\pm$0.004, respectively.
Izotov et al. (1997a) note that underlying stellar absorption 
in SBS 0335--052 plays minor role. However, unlike I Zw 18, fluorescent 
enhancement is important in SBS 0335--052
and should be taken into account.

    One of the important questions is how robust are measurements
of the He abundance in BCGs. We discuss this problem
using new high signal-to-noise ratio observations of the two most
metal-deficient BCGs known, I Zw 18 and SBS 0335--052. 
Our main motivation for the current study is to derive the
helium mass fraction $Y$ in these
extremely metal-deficient BCGs with better precision and to compare with 
results obtained in previous papers. This comparison allows us to analyze the
systematic effects introduced by different methods of He abundance
determination. Because 
these galaxies have very low oxygen abundances ($Z_\odot$/50 and
$Z_\odot$/40 in I Zw 18 and SBS 0335--052 respectively) the helium mass fraction
in these galaxies should be very close to $Y_p$ which we derive
here as the mean of the $Y$s in the two galaxies, after a 
small correction for stellar helium production.
The observations and data reduction are described in \S2. In \S3 we discuss
the heavy element abundances in I Zw 18 and SBS 0335--052. We derive  
$Y_p$ and the baryon mass fraction $\Omega_b$
and place constraints on the number of light neutrinos
in \S4. The results are summarized in \S5.

\section {OBSERVATIONS AND DATA REDUCTION}

   Spectrophotometric observations of I Zw 18
were obtained with the Multiple Mirror Telescope (MMT) on the nights
of 1997 April 29 and 30. Observations were made with the blue channel of the 
MMT spectrograph using a highly optimized Loral 3072$\times$1024 CCD
detector. A 1\farcs5$\times$180\arcsec\ slit was used along with a 300 groove
mm$^{-1}$ grating in first order and an L-38 second-order blocking filter. 
This yields a spatial scale along the slit of 0\farcs3 pixel$^{-1}$, a scale
perpendicular to the slit of 1.9\AA\ pixel$^{-1}$, a spectral range of
3600 -- 7500\AA, and a spectral resolution of $\sim$ 7\AA\ (FWHM). For these
observations, CCD rows were binned by a factor of 2, yielding a final sampling
of 0\farcs6 pixel$^{-1}$. The observations cover the full spectral range in a
single frame that contains all the lines of interest and have sufficient 
spectral resolution to distinguish between narrow nebular and broad WR emission
lines. The total exposure time was 180 minutes and was broken up into six
sub-exposures of 30 minutes each. All exposures were taken at small air masses
(1.1 -- 1.2), so no correction was made for atmospheric dispersion. The seeing
 was 0\farcs7 FWHM. The slit was oriented in the direction
with position angle P.A. = --41\arcdeg\ to permit observations of both NW
and SE components. The spectrophotometric standard star HZ 44 was
observed for flux calibration. Spectra of He-Ne-Ar comparison lamps were
obtained after each subexposure to provide wavelength calibration.

The Keck II telescope optical spectra of SBS 0335--052 were obtained on
1998 February 24 with the Low Resolution Imaging Spectrometer (LRIS)
(Oke et al. 1995), using the
300 groove mm$^{-1}$ grating which provides a dispersion 
2.52 \AA\ pixel$^{-1}$ and a
spectral resolution of about 8 \AA\ in first order. The total spectral
range covered was $\lambda$$\lambda$3600--7500 \AA. The slit was 
1\arcsec$\times$180\arcsec\ and centered on the second brightest stellar
cluster (cluster No. 5 in Thuan, Izotov \& Lipovetsky 1997) in order to look 
for WR stellar emission. The slit was oriented with a position angle 
P.A. = 60\arcdeg\ perpendicular to the SBS 0335--052 major axis. No binning
along the spatial axis has been done, yielding a spatial sampling of
0\farcs2 pixel$^{-1}$. The total exposure time was 40 min, broken into two
exposures of 30 min and 10 min. Both exposures were taken at air mass
1.4, and no correction was made for atmospheric dispersion. 
The seeing was 0\farcs6. In the 30 min spectrum
the two strongest lines [O III] $\lambda$5007
and H$\alpha$ were saturated in the 5 central rows. Therefore, the [O III]
$\lambda$5007/H$\beta$ and H$\alpha$/H$\beta$ intensity ratios in each of the
five central rows in the second 10 min exposure were used to correct
for saturation. The standard star G191B2B
was observed for absolute flux calibration. The spectrum of a He--Ne--Ar 
comparison lamp was obtained after the observation to provide wavelength 
calibration.

The slit orientations during the observations of I Zw 18 and SBS 0335--052
superposed on archival {\sl Hubble Space Telescope} ({\sl HST})\footnote[3]{
Observations are obtained with NASA/ESA {\sl Hubble Space Telescope} through
the Space Telescope Science Institute, which is operated by AURA Inc., under
NASA contract NAS 5-26555.} WFPC2 $V$ images are shown in Figure 1.
For the MMT observations the slit crossed the C component of I Zw 18 seen 
in Figure 1 at the NW edge of the slit.
Recently, the spectral energy distribution of this component has been discussed 
by van Zee et al. (1998) and Izotov \& Thuan (1998b), and will not be commented
on further. We only note that the
brightest emission lines H$\beta$ and H$\alpha$ in this component 
are detected in the MMT spectrum
as well. In the case of SBS 0335--052 the slit crossed the H II region 
$\sim$ 1\farcs2 to the NW from the brightest knot perpendicular to the
direction used by Izotov et al. (1997a).
   
The data reduction was carried out at the NOAO headquarters in Tucson using
the IRAF\footnote[4]{IRAF: the Image Reduction and Analysis Facility is
distributed by the National Optical Astronomy Observatories, which is
operated by the Association of Universities for Research in Astronomy,
In. (AURA) under cooperative argeement with the National Science
Foundation (NSF).} software package. Procedures included bias subtraction,
cosmic-ray removal and flat-field correction using exposures of a
quartz incandescent lamp. After wavelength mapping and night sky background
subtraction each frame was corrected for atmospheric extinction and flux
calibrated.

   Different apertures have been used for the extraction
of one-dimensional spectra for both galaxies. 
In Figure 2a we show the one-dimensional spectrum 0\farcs6$\times$1\farcs5 
aperture of the brightest part of the NW component of I Zw 18. 
The spectrum shows broad WR bumps at 
$\lambda$4650 and
$\lambda$5808 which have been discussed by Izotov et al. (1997b) and Legrand et 
al. (1997). Figure 2b shows the spectrum in a 0\farcs6$\times$1\farcs5 aperture 
of the SE component at the angular distance of 5\farcs4 from the NW component. 
The location of the He I and He II lines is marked in both spectra. 
The spectra of the NW and SE components are quite different. All He I lines 
in the spectrum of the SE component are in emission while $\lambda$4026 and 
$\lambda$4921 are in absorption and $\lambda$4471 is barely seen in emission
in the spectrum of the NW component. To the best of 
our knowledge, this is the first direct detection of He I absorption
lines in the spectrum of I Zw 18. Three other He I lines marked in the spectrum
of the NW component are in emission, although their intensities are reduced
due to the presence of underlying stellar absorption. Additionally, the
intensity of the He I $\lambda$5876 emission line is affected by Galactic
neutral sodium absorption.

   To improve the signal-to-noise ratio for heavy element and helium abundance
determinations, we use one-dimensional MMT spectra of the NW and the SE 
components of I Zw 18 in 4\farcs2$\times$1\farcs5 and 
3\farcs5$\times$1\farcs5 apertures respectively, separated by 5\farcs4. 
The observed line intensities and those corrected for interstellar extinction 
and underlying stellar absorption are shown in Table 1 along with the equivalent 
width $EW$(H$\beta$), the observed 
flux of the H$\beta$ emission line, the extinction coefficient $C$(H$\beta$)
and the equivalent width of hydrogen absorption lines $EW$(abs). To 
correct for extinction we used the Galactic reddening law by Whitford 
(1958).

   Taking advantage of the very high signal-to-noise ratio of the Keck II 
spectrum of the brightest part of SBS 0335--052 (SNR $\geq$ 100 for the 
continuum) we have defined several apertures for extraction of 
one-dimensional spectra. In Figure 3 we show the spectrum in a 
0\farcs6$\times$1\farcs0 aperture of the brightest part. 
We report the discovery of a weak WR bump, seen only in the brightest part of 
SBS 0335--052 (Figure 4). The properties of the WR stellar population in 
SBS 0335--052 will be discussed in a later paper. Here we note only that Wolf-Rayet
stars have now been seen in the two most metal-deficient galaxies known, 
I Zw 18 and
SBS 0335--052. This fact puts strong constraints on massive star evolution 
models at very low metallicities.
We also extract eight other one-dimensional spectra with the same 
0\farcs6$\times$1\farcs0 aperture 
separated from one another by 0\farcs6. The observed and corrected
line intensities in spectra to the SW of the brightest part, called
0\farcs6SW, ..., 2\farcs4SW and in the spectra to the NE of the brightest
part (0\farcs6NE, ..., 2\farcs4NE) are presented in Table 2 together with line
intensities in the spectrum of the brightest part called ``Center''.

 The errors in the line intensities (Tables 1, 2) have been calculated 
from the noise statistics in the continuum which include implicitly the errors 
introduced during the data reduction process (flat-field correction, sky 
subtraction), those coming from the placement of 
the continuum and fitting the line profiles with gaussians. We also
take into account the errors introduced by uncertainties in the spectral energy
distributions of standard stars. Standard star flux deviations for both standard
stars G191B2B and HZ 44 are taken to be 1\% (Oke 1990; Bohlin 1996). 
These errors have been propagated in the calculations of element abundances.
The main contribution to element abundance errors in I Zw 18 and in the
brightest regions of SBS 0335--052 comes from uncertainties in flux calibration
while for other apertures the main source of uncertainty originates from errors 
in the measurements of line intensities.

\section{HEAVY ELEMENT ABUNDANCES}

%\subsection{Electron temperature and electron number density}

 To derive element abundances, we have followed the 
procedure detailed in ITL94 and ITL97. We adopt a two-zone photoionized H II 
region model (Stasi\'nska 1990): a high-ionization zone 
with temperature $T_e$(O 
III), and a low-ionization zone with temperature $T_e$(O II).  We have  
determined $T_e$(O III) from the 
[O III]$\lambda$4363/($\lambda$4959+$\lambda$5007) ratio 
using a five-level atom model. That temperature is used for the 
derivation of the He$^+$, He$^{2+}$, O$^{2+}$, Ne$^{2+}$ and Ar$^{3+}$ ionic 
abundances. To derive $T_e$(O II), we have used 
the relation between
$T_e$(O II) and $T_e$(O III) (ITL94), based on a fit to the
photoionization models of Stasi\'nska (1990). The temperature $T_e$(O II) is 
used to derive the O$^+$, N$^+$ and Fe$^+$ ionic abundances. For Ar$^{2+}$
and S$^{2+}$ we have used an electron temperature intermediate between
$T_e$(O III) and $T_e$(O II) following the prescriptions of Garnett (1992).
The electron density $N_e$(S II) is derived from
the [S II] $\lambda$6717/$\lambda$6731 ratio.
Total element abundances are derived after correction for unseen stages
of ionization as described in ITL94 and Thuan, Izotov \& Lipovetsky (1995).

\subsection{I Zw 18}

The values of the derived electron temperature and number density in the NW
and the SE components of I Zw 18 (Table 3) are in good agreement with those by 
SK93, except for the electron temperature in the SE component, which is 1900 K 
higher in our case. Our determination of $T_e$ and $N_e$ is in very good
agreement with the recent study of I Zw 18 by Izotov \& Thuan (1998b). The
low value of the electron number density in the SE component makes it a suitable
target for the determination of the helium abundance because
the collisional enhancement of He I emission lines 
plays a minor role in this case.

   The oxygen abundances derived in the NW and the SE components of I Zw 18,
12 + log O/H = 7.166$\pm$0.027 and 7.183$\pm$0.025, respectively, are in very
good agreement with the values 7.16$\pm$0.01 and 7.19$\pm$0.02 
derived by Izotov \& Thuan (1998b), suggesting the robustness of 
the oxygen abundance
determination and a nearly constant metallicity in I Zw 18 when measured in
large apertures. However, smaller spatial variations of oxygen abundance 
correlated with small scale variations of the electron temperature may be 
present. In Figure 5a we show the spatial distribution of the electron 
temperature $T_e$(O III) in I Zw 18 along the slit. We note a possible
gradient of the electron temperature in the NW component 
with a highest temperature of $\sim$ 21,500 K 
in the brightest part associated with the Wolf-Rayet stars. 
Small-scale spatial variations of the oxygen abundance are shown in Figure 5b
( 1\arcsec\ = 52 pc at the distance $d$ = 10.8
Mpc to I Zw 18). The lowest oxygen abundance 12 + log O/H = 7.07$\pm$0.06 is 
found in the center of the NW component. This value of the oxygen abundance is
derived in the region where WR stars are present and, hence, may be too low
because of possible temperature fluctuations on small scales.
The presence of small oxygen abundance variations is in agreement with
findings of Martin (1996) and V\'ilchez \& Iglesias-P\'aramo (1998).
However, the variations of the electron temperature and oxygen abundance in 
I Zw 18 are only marginally statistically significant 
because of the large errors. The other heavy element abundances derived in 
large apertures are in good agreement with those derived by Izotov \& Thuan (1998b).

\subsection{SBS 0335--052}

   The spatial variations of the electron temperature $T_e$(O III) and oxygen
abundance 12 + log (O/H) in SBS 0335--052 are shown in Figures 5d and 5e
respectively. At distances $>$ 3\arcsec\ from the center of SBS 0335--052
the determinations of $T_e$(O III) and 12 + log (O/H) are subject to large
uncertainties, therefore we consider only the central part with size
$\leq$ 6\arcsec\ or $\sim$ 1600 pc ( 1\arcsec\ = 262 pc at the distance $d$ = 54.1
Mpc to SBS 0335--052). The lowest electron temperature $T_e$(O III) 
of $\sim$ 20,000 K is
found in the brightest part of the supergiant H II region. It is among
the highest temperatures found for H II regions in BCGs, reflecting the very low 
metallicity in SBS 0335--052. At larger distances from the brightest part, 
the temperature reaches values as high as $\sim$ 22,000 K. Such high values of 
$T_e$(O III) in SBS 0335--052 have also been found by Izotov et al. (1997a) 
in directions perpendicular to those studied here. 
We expect these variations to be real because of the small errors 
in the electron temperature determination. This suggests that the
electron temperature in the central part is statistically different from 
the higher temperatures derived in the regions to the SW of the central 
part. These temperature differences are
associated with differences in oxygen abundance; a statistically
significant gradient of oxygen abundance is evidently present in Figure 5e.
Izotov et al. (1997a) have also noted a gradient of oxygen abundance
along another direction. The highest oxygen abundance 12 + log (O/H) = 
7.338$\pm$0.012 in the region 0\farcs6NE is $\sim$ 1.4 times larger than
the oxygen abundance of $\sim$ 7.2 derived in the SW part of the H II region.
This gradient might be due to oxygen enrichment 
by the stellar clusters in the central part of the galaxy.
While the transport to large distances and mixing of freshly synthesized heavy 
elements may be efficient due to the presence of high-speed gas motion within 
the H II region (Izotov et al. 1997a), Figure 5e tells us that this process is 
incomplete in SBS 0335--052.

   In Table 4 we show $T_e$, $N_e$ and heavy element abundances for SBS 0335--052 
derived in 9 equal apertures but in different regions along the slit.
In contrast to I Zw 18, the electron number density $N_e$(S II) in SBS 0335--052 
is high. Large $N_e$(S II) values in this galaxy have also been derived 
by Melnick, Heydari-Malayeri \& Leisy (1992), Izotov et al.
(1997a) and Izotov \& Thuan (1998a).
Note the slight increase of electron temperature and the decrease of electron
number density with distance from the center of the H II region. The 
other heavy-element-to-oxygen abundance ratios are constant
within the errors.

   Most puzzling is the iron abundance in SBS 0335--052. While [O/Fe]
([O/Fe] $\equiv$ log (O/Fe) -- log (O/Fe)$_\odot$) in other BCGs 
is at the level 0.3 -- 0.5 (Thuan, Izotov \& Lipovetsky 1995; Izotov \& Thuan 
1999), suggesting a significant overproduction of oxygen relative to iron, similar 
to that observed in halo stars, [O/Fe] in SBS 0335--052 is $\sim$ 0. 
Izotov et al. (1997a) first noted this as atypical for BCGs. Izotov \& Thuan 
(1998a) later confirmed the low value. The reason for this is unclear. 
Izotov \& Thuan (1999) suggested that the low value of 
[O/Fe] is probably caused by the contamination and hence artificial
enhancement of the nebular [Fe III] $\lambda$4658 emission line by the
stellar C IV $\lambda$4658 emission line produced in hot stars with 
stellar winds. Independent of the problem of the origin of iron in SBS 
0335--052, our analysis shows that
heavy element abundances in I Zw 18 and SBS 0335--052, derived by different
authors from different observations, are very consistent.
This testifies to the robustness of element abundance determination in BCGs.

\section{HELIUM ABUNDANCE}
   He emission-line strengths are converted to singly ionized helium 
$y^+$ $\equiv$ He$^+$/H$^+$ and doubly ionized helium $y^{++}$ $\equiv$ 
He$^{++}$/H$^+$
abundances using the theoretical He I recombination line emissivities 
by Smits (1996). The analytical fits to Smits' emissivities given by ITL97 
agree with those by Brocklehurst (1972) for 
the He I $\lambda$4471 line within $<$1.5\%, for the He I $\lambda$5876 line
within $<$0.3\% and for the He I $\lambda$6678 line within $<$0.4\% in 
the temperature range from 10,000 to 20,000K. However, Smits' (1996) 
emissivity for the He I $\lambda$7065 line is $\sim$ 40 percent higher than 
the incorrect Brocklehurst value in the same temperature range. From this 
comparison the systematic error in helium abundance 
determination due to errors in the
He I emissivities is expected to be less than 0.5\% because of the high weight
of the strongest He I $\lambda$5876 emission line 
(see also Peimbert 1996). Recently Benjamin, Skillman
\& Smits (1999) have derived new He I emissivities that include the effects of
collisional excitation from the 2$^3$S metastable level. They estimate that
atomic data uncertainties alone may limit abundance estimates to an accuracy
of $\sim$ 1.5\% which is $\sim$ 3 times worse than the accuracy 
expected from a simple comparison of Brocklehurst (1972) and Smits (1996)
emissivities. Benjamin et al. (1999) note, however, that their error estimates 
of emission line intensities are somewhat
subjective and should be considered illustrative rather than definitive.
We therefore do not take into account the uncertainties in atomic data
in our error estimates for $^4$He abundance determination.

   To obtain the total helium abundance, the fraction of unseen neutral helium 
needs to be considered. For H II regions with  
ionizing stars having effective temperatures larger than 40000 K,
the He$^+$ and H$^+$ zones are mostly coincident (Stasi\'nska 1990).
To estimate the correction factor for neutral helium, we have  
used the ``radiation softness parameter'' $\eta$ of V\'ilchez \& Pagel (1988)
\begin{equation}
\eta = \frac{{\rm O}^+}{{\rm S}^+}\frac{{\rm S}^{++}}{{\rm O}^{++}}.       
\label{eq:eta}
\end{equation}
The fraction of neutral helium is significant ($\geq$ 5\%) when 
$\eta \geq$ 10 (Pagel et al. 1992). In our galaxies, the value of $\eta$ less 
than 1.5. Thus, the contribution of neutral helium is very small ($<$ 
1\%) for our galaxies and we have chosen not to correct for it.

   Additionally, in those cases when the nebular He II $\lambda$4686 emission
line was detected, which is the situation for both I Zw 18 and SBS 0335--052,
we have added the abundance of doubly ionized helium $y^{++}$ 
to $y^+$. The value of $y$$^{++}$ lies in the range 1 -- 5\% of $y^+$ 
in I Zw 18 and SBS 0335--052.

   Finally the helium mass fraction is calculated by
\begin{equation}
Y=\frac{4y[1-20({\rm O/H})]}{1+4y},                     \label{eq:Y}
\end{equation}
where $y$ = $y^+$ + $y^{++}$ is the number density of helium relative to 
hydrogen (Pagel et al. 1992).

The main mechanisms causing deviations of the He I emission line intensities 
from the recombination values are collisional and fluorescent enhancements.     
In order to correct for these effects, 
we have adopted the following procedure,  
discussed in more detail in ITL94 and ITL97: using the formulae by Kingdon \& 
Ferland (1995) for collisional enhancement and the Izotov \& Thuan 
(1998a) fits to Robbins (1968) calculations for fluorescent enhancement, we 
have evaluated the electron number density $N_e$(He II) and the optical depth 
$\tau$($\lambda$3889) in the He I $\lambda$3889 line in a self-consistent 
way, so that the He I $\lambda$3889/$\lambda$5876, 
$\lambda$4471/$\lambda$5876, 
$\lambda$6678/$\lambda$5876 and $\lambda$7065/$\lambda$5876 line ratios 
have their recombination values, after correction for collisional and 
fluorescent enhancement. The He I $\lambda$3889 and $\lambda$7065 lines play 
an 
important role because they are particularly sensitive to both optical depth
and electron number density. Since the
He I $\lambda$3889 line is blended with the H8 $\lambda$3889 line, we have 
subtracted the latter, assuming its intensity to be equal to 0.106 
$I$(H$\beta$) (Aller 1984),
after correction for the interstellar extinction and underlying
stellar absorption in hydrogen lines.
We adopt $N_e$=10 cm$^{-3}$ for the minimum number density in 
the calculations of collisional correction factors instead of the commonly 
used value of 100 cm$^{-3}$. With $N_e$=10 cm$^{-3}$, the collisional 
enhancement of the collisionally
sensitive He I $\lambda$5876 emission line is negligible, even
in the hottest H II regions of I Zw 18 and SBS 0335--052, 
while for $N_e$=100 cm$^{-3}$ the 
collisional enhancement can be as large as $\sim$ 5\%.
The singly ionized helium abundance $y^+$ 
and He mass fraction $Y$ is obtained for each of the 
three He I $\lambda$4471, $\lambda$5876 and $\lambda$6678 lines
by the self-consistent procedure. 
We then derived the weighted mean $y^+$ of these three determinations, the
weight of each line being determined by its intensity. However,
this weighted mean value may be underestimated due to the lower
value of $y^+$($\lambda$4471) resulting from underlying stellar absorption. 
Therefore, in subsequent discussions
we also use the weighted mean values of $Y$ derived from the
intensities of only two lines, He I $\lambda$5876 and $\lambda$6678.

Since the number density $N_e$(S II) is used to correct He I emission line 
intensities for collisional enhancement in many studies (e.g. in OSS97), we also 
consider this approach to study its validity in the
He abundance determination from our high quality observations. 

\subsection{Spatial distributions of He I and He II line equivalent widths 
and intensities in I Zw 18 and SBS 0335--052}

   Ideally, if the He abundance in an H II region is constant and no effect
is present to modify the He I emission line intensities, the latter
should be nearly the same at each point of the H II region. In reality, because 
the temperature in the H II region is not constant and because some He 
enrichment may be present in the central
parts, we may expect some variations of He I and He II
emission line intensities. However, these effects are small
due to the weak dependence of He I and He II recombination emissivities
on temperature and the small difference in oxygen abundance observed in
the central and outer parts of the H II regions in I Zw 18 and SBS 0335--052.
Therefore, if significant deviations of relative line intensities
from a constant value are observed in the spatial distribution of He I emission
lines, then some other mechanisms must be at work.

   Several mechanisms may play a role in changing the He I emission line
intensities from their recombination values. The efficiency of each mechanism is 
dependent on the particular physical conditions in the H II region. 
All these mechanisms need to be taken into account and several lines need to
be considered to discriminate between different mechanisms. 

    We already have shown that the physical conditions in I Zw 18 and SBS 
0335--052 are quite different. While in I Zw 18 the electron number density is 
small ($N_e$(S II) $\leq$ 100 cm$^{-3}$, Table 3) and collisional enhancement 
has a minor effect on the derived helium abundance, the electron number density 
in SBS 0335--052
is much higher ($N_e$(S II) $\sim$ 500 cm$^{-3}$ in the central part of the H II 
region, Table 4). Additionally, the linear size of the H II 
region in SBS 0335--052 is $\sim$ 5 times larger than in I Zw 18, and it may be 
optically thick for some He I transitions. 
Therefore, we expect both collisional and 
fluorescent enhancements of He I emission lines to play a significant role in 
this galaxy. By contrast, underlying stellar He I absorption is
much more important in I Zw 18 as is evident from Figures 5c and 5f where
the spatial distribution of the H$\beta$ emission line equivalent width
in both galaxies is shown. The equivalent widths of the He I emission lines
scale roughly as the equivalent widths of the H$\beta$ emission line. The 
smaller the equivalent width of the He I emission line is, the larger is the 
effect of underlying absorption.
In SBS 0335--052 the equivalent width of H$\beta$ is
high everywhere along the slit with a lowest value of 160\AA\ in the center
of the H II region and increasing to $\sim$ 300\AA\ in the outer parts.
In I Zw 18 the equivalent width of H$\beta$ emission line is much lower
with a minimum value of only 34\AA\ in the center of the NW component
or $\sim$ 5 times lower than in the center of SBS 0335--052. This is the
lowest value ever measured for $EW$(H$\beta$) in I Zw 18. 
Given equal equivalent widths for
He I absorption lines in both BCGs, we may expect the effect
of underlying stellar absorption to be several times larger in I Zw 18 than
in SBS 0335--052. The H$\beta$ equivalent width in the SE component is 
as high as $\sim$ 150\AA\ and therefore the effect of underlying stellar 
absorption, although present, is several times smaller. This is clearly seen
in Figures 2a and 2b. While higher order Balmer hydrogen lines and some helium
lines in the spectrum of the NW component are in absorption, all these lines
are in emission in the spectrum of the SE component. Even a visual inspection
of the NW component spectrum is sufficient to reveal that this component is not 
suitable for He abundance determination.

    In Figure 6 we show the spatial distribution of the He I $\lambda$4471, 
$\lambda$5876, $\lambda$6678 and $\lambda$7065 emission line equivalent widths 
in I Zw 18 and SBS 0335--052 along the slits. The minimum equivalent 
widths of He I lines are spatially coincident with the maxima in the continuum
intensity distribution. They are indicated in Figure 6 by their values. 
There is a significant difference between I Zw 18 and SBS 0335--052. While the 
maximum equivalent widths of the He I emission lines
in the SE component of I Zw 18 are close to those in the central brightest 
part of SBS 0335--052, the He I line equivalent widths in the NW component of
I Zw 18 are several times smaller. The largest ratio of minimum
values of equivalent widths in I Zw 18 as compared to in SBS 0335--052 
is found to be $\sim$ 14 for the He I $\lambda$4471 emission line. This implies
that He I $\lambda$4471 emission line intensity in the center of the NW 
component of I Zw 18 is affected by underlying stellar absorption which is an 
one order of magnitude larger as compared to that in the center of SBS 
0335--052. The effect of underlying stellar absorption on the He I $\lambda$6678 
emission line is smaller
due to its higher equivalent width. It follows from Figure 6 that  
influence of underlying stellar absorption on the intensity of He I
$\lambda$6678 emission line is much larger in the NW component of I Zw 18
than in its SE component and in SBS 0335--052.

    He I absorption lines are observed in the spectra of O and B stars with 
maximum values of equivalent widths occuring in B2 -- B3 stars (Jaschek et al.
1994; Leone \& Lanzafame 1998). Olofsson (1995)
has produced spectral evolutionary models of single-burst star forming regions
in order to follow the behaviour of hydrogen and helium absorption lines.
Using these calculations we can estimate the effect of underlying stellar
absorption on the He I emission line intensities in our galaxies. Olofsson (1995)
found that typical equivalent widths of the He I $\lambda$4471 absorption line
in starbursts with age $\leq$ 4 -- 6 Myr lie in the range 0.2 -- 0.3\AA.
Therefore, underlying absorption decreases the intensity of the 
He I $\lambda$4471 emission line in the center 
of the NW component of I Zw 18 by factors up to $\sim$ 2. 
This effect on the He I $\lambda$4471 emission
line intensity in the central
part of SBS 0335--052 is smaller, but it is significant and the correction for
underlying stellar absorption is $\sim$ 5 -- 7\%. 
Olofsson (1995) does not provide calculations for the He I $\lambda$5876, 
$\lambda$6678 and $\lambda$7065 absorption lines. The equivalent widths of 
these lines in OB stars are respectively $\sim$ 2, $\sim$ 2 and $\sim$ 3
times smaller than that for the He I $\lambda$4471 absorption line
(Leone \& Lanzafame 1998). Additionally,
the equivalent width of absorption lines in the red part of star-forming region
spectra is diluted by the larger contribution of cooler stars and gaseous
emission (Izotov et al. 1997a; Papaderos et al. 1998). Therefore, we put a
conservative upper limit of 0.1\AA\ for the equivalent widths of the He I
$\lambda$5876, $\lambda$6678 and $\lambda$7065 absorption lines. The correction
of the He I $\lambda$6678 emission line for underlying absorption in the NW
component is as high as 5\% and must be taken into account. 
On the other hand, the
correction of the same line intensity in the SE component of I Zw 18 and 
in SBS 0335--052 is $\leq$1\%, while the correction of He I $\lambda$5876
emission line is only $\leq$0.4\%. 

This comparison of H II region properties in the two most metal-deficient
BCGs, I Zw 18 and SBS 0335--052, shows that these galaxies are well-suited
for studying all effects influencing He line intensities.
In Figure 7 we show the spatial distribution of He I and He II $\lambda$4686 
emission line intensities in both BCGs. A decrease of the
He I $\lambda$4471, $\lambda$5876 and $\lambda$6678 emission intensities is
evident in the NW component of I Zw 18. Their intensities are 
smaller by factors of 
$\sim$ 3, $\sim$ 2 and $\sim$ 1.2 respectively compared to the 
intensities in the SE component. Only $\sim$ 5\% of this decrease can be 
explained by the presence of doubly ionized He. The rest of
the He I intensity decrease is evidently due to the underlying stellar He I
absorption lines. 

   The spatial distribution of He I emission line intensities in SBS 0335--052 
is very different from that in I Zw 18. The increase of He I $\lambda$5876 and 
He I $\lambda$7065 emission line strengths by $\sim$ 20\% and $\sim$ 75\% respectively
in the central part of SBS 0335--052 within a radius $\sim$ 2\arcsec\ is
caused by collisional and fluorescent enhancement. The increase
of the He I $\lambda$6678 emission line intensity is only $\leq$ 4\%.
The combined effect of collisional enhancement and underlying stellar absorption
results in a small depression in the He I $\lambda$4471 
intensity in the central region. 
The intensity of this line is also affected by underlying absorption.  However, 
this effect is significantly smaller than that in the NW component of I Zw 18.

\subsection{Helium mass fraction in I Zw 18}

    In Table 5, the results of the He mass fraction determination are given for 
both the NW and the SE components. We calculate the He mass fraction in
the NW component to show the severity of the effects of underlying stellar
absorption and of absorption by Galactic interstellar sodium on the final 
results. The He mass fractions in the NW and the 
SE components should be the
same. However, the calculated He mass fractions $Y$($\lambda$4471) = 
0.169$\pm$0.023 and $Y$($\lambda$5876) = 0.192$\pm$0.007 in the NW component are 
unreasonably small and far below the values derived for the SE component.
Although the He mass fraction $Y$($\lambda$6678) = 0.237$\pm$0.017 is much higher, 
it is also subjected to underlying stellar absorption and should not be 
used without correction for this effect which can be as high as 5\%. 
To correct for 
underlying He I absorption detailed calculations of stellar population
synthesis models as a function of time are required because the equivalent 
widths of He I absorption lines are time-dependent, being 1.3 -- 1.5 times 
larger in O, B supergiants compared to 
those in O, B main sequence stars (Jaschek et al. 1994). As mentioned before,
although the calculations of some absorption
He I equivalent widths in young stellar clusters are now available 
(e.g. Olofsson 1995), such calculations have not been made for the He I
$\lambda$5876, $\lambda$6678 and $\lambda$7065 absorption lines.

    The He mass fractions $Y$($\lambda$5876) = 0.243$\pm$0.008 and
$Y$($\lambda$6678) = 0.243$\pm$0.018 derived in the SE component are in
very good agreement. The $Y$($\lambda$4471) = 0.232$\pm$0.018 is lower
due to underlying stellar absorption. The weighted mean values 
$Y$ = 0.2416$\pm$0.0065 (three lines) and 0.2429$\pm$0.0070 ($\lambda$5876 +
$\lambda$6678) 
in the SE component are in excellent agreement with value $Y$ = 0.242$\pm$0.009
derived by Izotov \&
Thuan (1998b), much better than is expected from the calculated errors.
The result is not dependent on the method used because the electron 
number density in the SE component is so small that the correction for collisional 
enhancement is $\leq$ 0.5\% for the He I $\lambda$5876 emission 
line which is most subject to that effect. Therefore, we conclude that the SE 
component of I Zw 18 is an excellent target for a more precise determination of 
the He abundance with 10-meter class telescopes, as the precision in $Y$ for 
this component is limited mainly by observational uncertainties.

\subsection{Helium mass fraction in SBS 0335--052}

    Correction of the He I emission lines for collisional and fluorescent
enhancements in SBS 0335--052 presents more problems due to its higher
electron number density and the larger extent of the H II region as compared
to the SE component of I Zw 18. These problems have been discussed by
Izotov et al. (1997a) and Izotov \& Thuan (1998a). 
The importance of the fluorescent mechanism has been demonstrated by Osterbrock,
Tran \& Veilleux (1992) and Esteban et al. (1998) for the Orion nebula and
by Thuan, Izotov \& Lipovetsky (1996) for the BCG Mrk 996, where this effect is
especially strong due to the extremely high electron number density 
($N_e$ $\sim$ 10$^6$cm$^{-3}$) in its H II region. Although 
much smaller, fluorescent enhancement is present in SBS 0335--052 as well
(Izotov et al. 1997a).

    The very high signal-to-noise ratio spectrum of SBS 0335--052 obtained with
the Keck II telescope allows us to measure the He mass fraction
in several regions of SBS 0335--052 and to significantly improve
the measurement precision. 

    The results of the He abundance determination in 9 different regions of
SBS 0335--052 are shown in Table 6, both for the self-consistent
method and the method using $N_e$(S II). 

Consider first the results obtained
with the self-consistent method. The electron number density $N_e$(He II)
derived by this method is significantly lower than that derived from the
[S II] emission line ratio. Therefore, in the self-consistent method, 
the correction for collisional enhancement is smaller. Inspection of 
Table 6 shows the
remarkable coincidence of He mass fractions derived from He I $\lambda$5876
and He I $\lambda$6678 emission lines in each region, 
better than expected from the formal errors. 
By contrast, the He mass fraction derived from He I $\lambda$4471 is 
systematically lower because this line is subject to underlying stellar absorption
in the central brightest part of the galaxy with lowest $EW$(H$\beta$).
Additionally, we cannot completely exclude uncertainties in 
atomic data for this line (ITL97, OSS97). This effect is also seen in Figure 8a,  
where error bars are shown only for $Y$s derived from the He I $\lambda$5876 
emission line. Note the constancy and coincidence of $Y$ values derived from the 
He I $\lambda$5876 and He I $\lambda$6678
emission lines. The weighted mean He mass fractions for 9 different regions
are $Y$($\lambda$5876) = 0.2462$\pm$0.0018, $Y$($\lambda$6678) = 
0.2464$\pm$0.0029, $Y$($\lambda$5876 + $\lambda$6678) =
0.2463$\pm$0.0015 and $Y$(all lines) = 0.2437$\pm$0.0014. These values are
in excellent agreement with $Y$ = 0.245$\pm$0.006 derived from MMT 
observations (Izotov et al. 1997a),
although the errors in the present determination are several times smaller.
This result shows how observations of low-metallicity galaxies with the largest 
telescopes will allow us to improve the precision in $Y$ determination.

    The situation is very different when He I emission lines are corrected only
for collisional enhancement with the electron number density $N_e$(S II)
derived from [S II] emission lines. Inspection of Table 6 shows
that $Y$ values derived from the He I $\lambda$5876 emission line 
are systematically lower compared to those derived from the He I $\lambda$6678 
emission line. This systematic effect is evidently present in Figure 8b, where
we show error bars only for $Y$s derived from the He I $\lambda$5876 emission
line. Note that these error bars are larger than those in 
Figure 8a because of the large uncertainties in the determination of the
electron number density from [S II] lines.  The weighted mean values from 9
different regions are $Y$($\lambda$5876) = 0.2361$\pm$0.0037, 
$Y$($\lambda$6678) = 0.2412$\pm$0.0032 and $Y$(all lines) = 0.2344$\pm$0.0021.
Although these values are formally consistent within 1$\sigma$, the consistency
of $Y$ values derived by the self-consistent method is much better. The main 
reason of the systematic differences between $Y$ derived from He I $\lambda$5876
and $\lambda$6678 emission lines is the overcorrection of He I line intensities
for collisional enhancement due to too high an adopted electron number density.
This systematic effect leads us to conclude that the
electron number density derived from [S II] emission lines should not be used 
in general for the determination of He abundance. In some cases, such
as in SBS 0335--052, this can lead to an incorrect result, underestimating the
helium mass fraction in this galaxy by as much as 5 -- 10\%.

\subsection{Primordial He abundance and baryonic mass fraction of the Universe}

    Since I Zw 18 and SBS 0335--052 are very metal-deficient galaxies
with oxygen abundances $\sim$ 50 and $\sim$ 40 times smaller than the
oxygen abundance in the Sun, their helium mass fractions should be very close to
the primordial value $Y_p$. The contribution of stellar He in I Zw 18 and 
SBS 0335--052 is only $\sim$ 0.0004 if He is produced by massive stars only
(with d$Y$/d$Z$ $\sim$ 1, Maeder 1992; Timmes, Woosley \& Weaver 1995). If the
empirical slope, d$Y$/d$Z$ = 2.4, derived by Izotov \& Thuan (1998a) 
for a sample of low-metallicity BCGs is adopted, then the stellar contribution 
of He is $\sim$ 0.0010 or $<$ 0.5\% of the primordial helium mass fraction. 
The primordial mass fractions derived from the weighted mean He mass fractions
in I Zw 18 and in SBS 0335--052 are $Y_p$ = 0.2426$\pm$0.0013 when all three
lines are taken into account, and $Y_p$ = 0.2452$\pm$0.0015 when only He I
$\lambda$5876 and $\lambda$6678 are considered. Between these determinations, 
the latter value is preferable as it is less subject to the effect of underlying 
absorption in the He I $\lambda$4471 emission line. This value is in excellent
agreement with the primordial He mass fraction $Y_p$ = 0.2443$\pm$0.0015 derived
by Izotov \& Thuan (1998a) from the $Y$ -- O/H linear regression for a sample
of 45 BCGs. Thus determinations of $Y_p$, be it by the linear regression
technique or by measuring the helium mass fraction in the most metal-deficient
BCGs, give nearly the same result. The concordance of $Y_p$ values derived in 
several papers (ITL97, Izotov \& Thuan 1998ab, this paper) is testimony to the 
robustness and reliability of the analysis techniques we have used. 

    We now discuss the cosmological implications of our $Y_p$ determination. 
In Figure 9 
solid lines show the primordial abundances of $^4$He, D, $^3$He and $^7$Li
predicted by standard big bang nucleosynthesis theory as a function
of the baryon-to-photon number ratio $\eta$. The dashed lines are 1$\sigma$
uncertainties in model calculations. 
The analytical fits are taken from Fiorentini et al. (1998). The solid boxes 
show the 1$\sigma$ predictions of $\eta$ as inferred from our measured primordial 
abundance of $^4$He, and the primordial abundances of D (Levshakov et al. 1999; 
Burles \& Tytler 1998b), $^3$He (Rood et al. 1998) and $^7$Li
(Bonifacio \& Molaro 1997; Vauclair \& Charbonnel 1998; Pinsonneault
et al. 1998). All these determinations are consistent to within 1$\sigma$. The 
uncertainties in the $^3$He abundance
are still large however, and do not provide as yet a stringent constraint of
the baryonic mass fraction. For comparison, we have also shown with 
dotted boxes the 1$\sigma$ predictions of $\eta$ inferred from the primordial
abundances of $^4$He derived by OSS97, of D (Songaila et al. 1997) and $^7$Li
(Bonifacio \& Molaro 1997; Vauclair \& Charbonnel 1998; Pinsonneault
et al. 1998). This second data set has, however, several features that
make it less than ideal for constraining $\eta$. 
OSS97 have included in their sample some low-metallicity galaxies with low He
abundances, including the NW component of I Zw 18. As shown here and by ITL97 
and Izotov \& Thuan (1998b) the He I emission line intensities in the NW 
component are strongly reduced by underlying stellar absorption and 
absorption by interstellar sodium. Therefore, this object
{\it must} be excluded from consideration which would result in a higher 
value of
$Y_p$ for the OSS97 sample. The helium mass fraction in the SE
component is in the range $Y$ = 0.242 -- 0.243 as shown by several 
independent studies by at least two different groups (Izotov \& Thuan 1998b; 
V\'ilchez \& Iglesias-P\'aramo 1998). Yet OSS97 derive $Y_p$ = 0.234.
It is most unlikely that $Y_p$ would be so much lower than the $^4$He mass 
fraction derived in the most metal-deficient galaxy known. Further, our 
analysis shows that neglecting fluorescent enhancement may 
artificially lower the helium mass fraction. The data for the 
whole OSS97 sample are 
not published, so we cannot check directly their quality and how they were
analyzed. However, some of their data for the most 
metal-deficient galaxies are taken from the literature and they were 
obtained many years ago with non-linear detectors. For instance, the 
BCG Tol 65 with a low oxygen abundance and a low $Y$ value has been observed in
the early 80s by Kunth \& Sargent (1983) and by Campbell, Terlevich \&
Melnick (1986) and should be reobserved for a more precise He abundance 
determination. 

As for deuterium, Levshakov et al. (1998, 1999) have redetermined the D 
abundance in high-redshift absorbing clouds. They have shown that
the use of more adequate kinematic models with correlated turbulent 
motions for absorbing Ly$\alpha$ clouds instead of microturbulent models
results in a narrow range of (D/H)$_p$ = (4.35$\pm$0.43)$\times$10$^{-5}$. 
Although these values are close to (D/H)$_p$ = (3.3$\pm$0.3)$\times$10$^{-5}$ 
and (3.39$\pm$0.25)$\times$10$^{-5}$ derived by 
Burles \& Tytler (1998ab), they are systematically higher
and are more consistent with our $^4$He abundance
determination (ITL97; Izotov \& Thuan 1998a; this paper). 
The high primordial D abundance reported by Songaila et al. (1997)
is probably excluded by this analysis. However, problems with the
interpretation of spectra of absorbing gas clouds and the determination of
D/H remain (e.g. Tytler et al. 1999) and measurements of D/H along more
lines of sight are needed.

   The primordial $^7$Li abundance determination suffers from uncertainties
arising from the possible depletion of this element inside stars. If this
depletion is present, then the primordial $^7$Li abundance is higher than the
value ($^7$Li/H)$_p$ = (1.73$\pm$0.21)$\times$10$^{-10}$ obtained by Bonifacio 
\& Molaro (1997). Vauclair \&
Charbonnel (1998) have shown that the depletion of $^7$Li in Population II
stars may be as high as 30\%. They give for the primordial lithium
abundance ($^7$Li/H)$_p$ = (2.24$\pm$0.57)$\times$10$^{-10}$. 
Pinsonneault et al. (1998) have analyzed $^7$Li depletion in rotating stars. 
They found that the depletion factor can be as high as 1.5 -- 3 times, larger 
than the value obtained by Vauclair \& Charbonnel (1998). Pinsonneault et al. 
(1998) give a primordial lithium abundance ($^7$Li/H)$_p$ = 
(3.9$\pm$0.85)$\times$10$^{-10}$.

   The $Y_p$ = 0.2452$\pm$0.0015 value derived in this paper when only the two
He I lines $\lambda$5876 and $\lambda$6678 are used, implies a baryon-to-photon
number ratio $\eta$ = (4.7$^{+1.0}_{-0.8}$)$\times$10$^{-10}$. This 
ratio is $\eta$ = (4.0$^{+0.08}_{-0.06}$)$\times$10$^{-10}$ if
$Y_p$ = 0.2437$\pm$0.0014. These baryon-to-photon number ratios 
translate to baryon mass fractions $\Omega_b$$h^2_{50}$ = 
0.068$^{+0.015}_{-0.012}$ and 0.058$^{+0.010}_{-0.008}$ where $h_{50}$ 
is the Hubble constant in units of 50 
km s$^{-1}$Mpc$^{-1}$. Although both values of baryon mass fraction are 
consistent within the errors, we prefer the higher value because
it is based on a primordial He determination which is less subject to 
uncertainties caused by underlying He I stellar absorption.

    Our value of the baryon mass fraction is consistent with that
obtained by analysis of the
Ly$\alpha$ forest in a cold dark matter cosmology. Depending on the intensity
of diffuse UV radiation, the inferred lower limit is 
$\Omega$$h^2_{50}$ = 0.05 -- 0.10 ( Weinberg et al. 1997; Bi \& Davidsen 1997; 
Rauch et al. 1997), while Zhang et al. (1998) have derived 0.03 
$\leq$$\Omega$$h^2_{50}$ $\leq$ 0.08.

We use the statistical $\chi^2$ technique with the code described by
Fiorentini et al. (1998) and Lisi et al. (1999) to analyze the consistency of 
different sets of primordial $^4$He, D and $^7$Li abundances at the high end 
range of the baryon-to-photon number ratio, varying both $\eta$ and equivalent number 
of light neutrino species $N_\nu$. The lowest $\chi^2_{min}$ = 0.001 
results when $\eta$ = 4.45$\times$10$^{-10}$ and $N_\nu$ = 3.006 
for the set of primordial abundances with $Y_p$ = 0.2452$\pm$0.0015 (this paper), 
(D/H)$_p$ = 
(4.35$\pm$0.43)$\times$10$^{-5}$ (Levshakov et al. 1999) and ($^7$Li/H)$_p$ = 
(2.24$\pm$0.57)$\times$10$^{-10}$ (Vauclair \& Charbonnel 1998). Hereafter,
we shall designate this most consistent data set by B. The value
$\chi^2_{min}$ for other choices of primordial abundances is larger than 0.5,
being equal to 3.7 for the data set with 
$Y_p$ = 0.2452$\pm$0.0015 (this paper), (D/H)$_p$ = 
(3.39$\pm$0.25)$\times$10$^{-5}$ (Burles \& Tytler 1998b) and ($^7$Li/H)$_p$ = 
(1.73$\pm$0.21)$\times$10$^{-10}$ (Bonifacio \& Molaro 1998) (hereafter 
data set A).
 
The joint fits of $\eta$ and $N_\nu$ to data sets A and B are shown in
Figure 10a and 10b, respectively. 
The 1$\sigma$ ($\chi^2$ -- $\chi^2_{min}$ = 2.3) and
2$\sigma$ ($\chi^2$ -- $\chi^2_{min}$ = 6.2) deviations are shown 
respectively by thin and thick solid lines. We find the equivalent number of
light neutrino species for data set A to be in the range
$N_\nu$ = 2.9$\pm$0.3 (2$\sigma$) (Figure 10a), and for 
the most consistent data set B to be in the range $N_\nu$ = 3.0$\pm$0.3 
(2$\sigma$) (Figure 10b). Both values are consistent with the experimental 
value of 2.993$\pm$0.011 (Caso et al. 1998) shown by the dashed line. 
Therefore, we conclude that both data sets of primordial abundances of light 
elements are in good agreement with predictions of standard big bang 
nucleosynthesis theory. However, the agreement is better for data set B.

\section {SUMMARY AND CONCLUSIONS}

    In this study we have presented the results of very high signal-to-noise
ratio spectroscopy of the two most metal-deficient blue compact galaxies known,
I Zw 18 and SBS 0335--052, obtained with the Multiple Mirror Telescope and the 
Keck II telescope, respectively. Our main aim is to derive the primordial He 
mass fraction $Y_p$ based on the mean helium mass fractions $Y$ derived in each 
BCG.

We have obtained the following results:

  1. Assuming all systematic uncertainties are negligible, 
the weighted mean He mass fraction for the two most metal-deficient galaxies
I Zw 18 and SBS 0335--052 is $Y$ = 0.2436$\pm$0.0013 if three He I emission
lines are used and $Y$ = 0.2462$\pm$0.0015 if the He I $\lambda$4471
emission line is excluded. The primordial He mass fraction is derived after a
correction for the stellar contribution to the He mass fraction of
$\sim$ 0.0010 adopting an empirical slope d$Y$/d$Z$ = 2.4 from Izotov \& Thuan 
(1998a). We then determine the primordial He mass fractions
$Y_p$ = 0.2426$\pm$0.0013 and $Y_p$ = 0.2452$\pm$0.0015 when 
three and two He I emission lines are used respectively. 
These $Y_p$ values correspond to baryon-to-photon number ratios $\eta$ = 
(4.0$^{+0.8}_{-0.6}$)$\times$10$^{-10}$
and $\eta$ = (4.7$^{+1.0}_{-0.8}$)$\times$10$^{-10}$ which translate to baryon
mass fractions $\Omega_b$$h^2_{50}$ = 0.058$^{+0.010}_{-0.008}$ and
$\Omega_b$$h^2_{50}$ = 0.068$^{+0.015}_{-0.010}$. We prefer the latter value, as 
the other one is subject to uncertainties caused by underlying absorption in the 
He I $\lambda$4471 emission line.

   2. We performed a statistical $\chi^2$ analysis with the
code developed by Fiorentini et al. (1998)
and Lisi et al. (1999) to study the consistency of different sets of primordial
$^4$He, D and $^7$Li abundances, varying both the baryon-to-photon number ratio
$\eta$ and the equivalent number of light neutrino species $N_\nu$. The best
consistency is achieved for the set of primordial abundances $Y_p$ =
0.2452$\pm$0.0015 (this paper), (D/H)$_p$ = (4.35$\pm$0.35)$\times$10$^{-5}$
(Levshakov et al. 1999) and ($^7$Li/H)$_p$ = (2.24$\pm$0.57)$\times$10$^{-10}$
(Vauclair \& Charbonnel 1998) at $\eta$ = 4.45$\times$10$^{-10}$ and $N_\nu$ = 
3.006 with $\chi^2_{min}$ = 0.001. This set of data strongly supports
the standard big bang nucleosynthesis theory. We derive the equivalent number
of light neutrino species to be $N_\nu$ = 3.0$\pm$0.3 (2$\sigma$).

   3. We find direct evidence for the presence of underlying stellar
He I absorption lines in the spectrum of the NW component of I Zw 18. The
equivalent width of the H$\beta$ emission line in the very central part of the
NW component is very small, being only 34\AA, the smallest ever
measured in I Zw 18. The equivalent widths of the He I emission lines are
small, comparable to those for stellar He I absorption lines.
Therefore, the intensities of the strongest He I emission lines are 
significantly reduced by stellar He I absorption lines and the two weaker He I
$\lambda$4026 and $\lambda$4921 lines appear purely in absorption. 
This makes the use of the NW component of I Zw 18 for He abundance 
determination inappropriate.
This conclusion was also arrived at by ITL97 and Izotov \& Thuan (1998b).
In contrast, the influence of underlying stellar absorption on the intensities
of He I emission lines in the SE component of I Zw 18 is significantly smaller.
The weighted mean He mass fraction in the SE component is $Y$ = 0.241$\pm$0.007  
when the three He I lines, $\lambda$4471, $\lambda$5876 and $\lambda$6678, are taken
into account and $Y$ = 0.243$\pm$0.007 when the He I $\lambda$4471 emission line
is excluded. These values are in excellent 
agreement with $Y$ derived recently by Izotov \& Thuan (1998b) and V\'ilchez \&
Iglesias-P\'aramo (1998). The concordance
of different determinations of the He mass fraction from 
different observations of I Zw 18 by different groups shows their robustness. 
We find that collisional enhancement of the He I emission lines is very small 
despite the high electron temperature in the H II region of I Zw 18. 
Observational errors are presently the
main source of uncertainty in determining the He mass fraction in
the SE component. Future observations with 10-meter class telescopes can 
significantly improve the measurement precision of $Y$ in this galaxy.

   4. The very high signal-to-noise ratio Keck II observations of SBS 0335--052
allow us to derive the helium mass fraction with good precision in nine
different regions in this BCG. We find that both collisional and fluorescent
enhancements of He I emission lines are important in SBS 0335--052, a conclusion  
reached previously by
Izotov et al. (1997a). It is shown that the electron number density derived
from [S II] emission lines cannot be used, because it leads to an 
overcorrection of the He I emission line intensities for 
collisional enhancement by 5 
-- 10\%, and hence to an underestimation
of the He mass fraction. Using only the single He I $\lambda$6678 emission 
line intensity corrected for collisional enhancement with the electron
number density $N_e$(S II) (as was done e.g. by OSS97) would lead to an
erroneous $Y$ value. Instead, a thorough analysis of the possible mechanisms 
which make He I emission line intensities deviate from
their recombination values {\it must} be made for {\it each} BCG
and a self-consistent method using several
He I emission lines should be applied to correct for both collisional and 
fluorescent enhancements. The helium mass fractions $Y$ derived from He I
$\lambda$5876 and $\lambda$6678 emission line intensities are in excellent 
agreement in each of the nine regions while the He mass fraction derived from 
He I $\lambda$4471 emission line is systematically lower due to the presence of
underlying stellar absorption and, possibly, due to uncertainties
in atomic data. The weighted mean of the He mass fraction for
all 9 regions of SBS 0335--052 is $Y$ = 0.2437$\pm$0.0014 if all three He I 
$\lambda$4471, $\lambda$5876 and $\lambda$6678 emission lines are used.
If the He I $\lambda$4471 emission line is excluded because of its sensitivity 
to underlying stellar absorption, then the weighted mean He mass fraction
becomes $Y$ = 0.2463$\pm$0.0015, which is our recommended value. These values of
He mass fraction in SBS 0335--052 are in good agreement with those
derived by Izotov et al. (1997a) and Izotov \& Thuan (1998a).

   5. The oxygen abundances 12 + log (O/H) = 7.17$\pm$0.03 and 7.18$\pm$0.03 
derived respectively in the NW and SE components of I Zw 18 agree very well
with the oxygen abundances derived in these components by Izotov \&
Thuan (1998b). While both components have nearly the same oxygen abundance, 
variations of chemical composition might be present at small scales as 
suggested by the spatial distribution of the oxygen abundance. Alternatively,
these variations might be caused by temperature fluctuations on small scales.

   6. The oxygen abundance 12 + log (O/H) = 7.321$\pm$0.011 in the brightest
part of SBS 0335--052 is in excellent agreement with the 
oxygen abundance derived
by Izotov et al. (1997a) using MMT observations with another slit 
orientation. We find a statistically significant gradient of oxygen abundance.
It is highest in the brightest part of the H II region associated 
with ionizing clusters and declines to a value 12 + log(O/H) $\sim$ 7.2 in 
the outer parts of the supergiant H II region, comparable to that found in the 
western component of SBS 0335--052 (Lipovetsky et al. 1999) and in I Zw 18. 
The existence of an oxygen abundance gradient with another 
slit orientation was previously noticed by Izotov et al. (1997a).

   7. We report the discovery of a WR stellar population in the BCG SBS 0335--052.
 The presence of WR stars in SBS 0335--052 was expected since
the UV {\sl HST} GHRS observations by Thuan \& Izotov (1997) showed stellar
Si IV $\lambda$1394 and $\lambda$1403 lines with P Cygni profiles suggestive
of the presence of hot massive stars with stellar winds. 
Recently Izotov et al. (1997b) and Legrand et al. (1997) have detected a WR
stellar population in the NW component of I Zw 18 as well. Thus, the two most 
metal-deficient BCGs  both contain WR stars. These findings are of special 
importance for constraining massive stellar evolution models at very 
low metallicity.

\acknowledgements
Y.I.I. thanks the staff of NOAO and the University of Virginia
for their kind hospitality. 
This international collaboration was possible thanks to the  
partial financial support of INTAS grant No. 97-0033
for which Y.I.I. and N.G.G. are grateful. C.B.F. acknowledges the
support of NSF grant AST-9803072. T.X.T. and Y.I.I. thank the partial financial
support of NSF grant AST-9616863.

%------------------------------------------------------------

\clearpage

\figcaption[fig1.ps]{ Slit orientations superposed on {\sl HST}
archival $V$ images of I Zw 18 and SBS 0335--052. The slit orientation
during the MMT observations of I Zw 18 is chosen in such a way as to get spectra
of the SE and NW components as well as of the C component to the NW of the
main body of the galaxy. The slit orientation during the Keck II observations
of SBS 0335--052 is chosen in the direction perpendicular to that used
by Izotov et al. (1997a) and it is centered on stellar cluster No. 5
(Thuan, Izotov \& Lipovetsky 1997). 
The spatial scale is 1\arcsec\ = 49 pc in the case of I Zw 18 and is
1\arcsec\ = 257 pc in the case of SBS 0335--052.}

\figcaption[fig2.ps]{The 0\farcs6$\times$1\farcs5 aperture MMT spectra
of the brightest parts of the NW and the SE components of I Zw 18. The spectrum
of the SE component is extracted at the angular distance of 5\farcs4 from
the NW component. The positions of He I lines
are marked. Note that all marked He I lines in the spectrum of the SE
component are in emission while the two He I $\lambda$4026 and $\lambda$4921
lines are in absorption and the He I $\lambda$4471 emission line is 
barely detected in the spectrum of the NW component.}

\figcaption[fig3.ps]{ The 0\farcs6$\times$1\farcs0 aperture Keck II 
LRIS telescope
spectrum of the brightest part of the SBS 0335--052. The positions of the He II
$\lambda$4686 and He I emission lines are marked.}

\figcaption[fig.4.ps] {Keck II LRIS spectrum of the brightest part of SBS 
0335--052 within a 0\farcs8$\times$1\farcs0 aperture.
The strongest nebular emission lines are marked. Note the presence of the
broad WR bump at rest wavelength $\lambda$4620 -- 4640\AA. The redshift of
SBS 0335--052 is $z$ = 0.0136. }

\figcaption[fig5.ps] {The spatial distributions of the electron temperature
$T_e$(O III), oxygen abundance 12 + log (O/H), and the equivalent width 
$EW$(H$\beta$) of the H$\beta$ emission line in I Zw 18 (left panel) and in
SBS 0335--052 (right panel). The error bars are 1$\sigma$ deviations.}

\figcaption[fig6.ps] {The spatial distributions of the He I nebular emission 
line equivalent widths in I Zw 18 (left panel) and in SBS 0335--052 (right 
panel). The error bars are 1$\sigma$ deviations. The value of the minimum 
equivalent width for each He I emission line is given.}

\figcaption[fig7.ps] {The spatial distributions of the He I and He II 
$\lambda$4686 nebular emission line intensities in I Zw 18 
(left panel) and in SBS 0335--052 (right panel). The error bars are 1$\sigma$ 
deviations.}

\figcaption[fig8.ps] {The spatial distributions of the helium mass fractions
in SBS 0335--052 derived from the He I $\lambda$4471, $\lambda$5876
and $\lambda$6678 emission line intensities. The intensities of the He I
emission lines in Figure 8a are corrected for fluorescent and collisional
enhancement with an electron number density $N_e$(He II) and an optical depth
$\tau$($\lambda$3889) derived self-consistently from the observed 
He I $\lambda$3889, $\lambda$4471, $\lambda$5876, $\lambda$6678 and 
$\lambda$7065 emission line intensities. The intensities of the He I emission
lines in Figure 8b are corrected only for collisional enhancement with an
electron number density $N_e$(S II). The 1$\sigma$ error bars are shown only for 
the He mass fraction derived from the He I $\lambda$5876 emission line. They are 
larger in Figure 8b because of the large uncertainties in the determination of 
$N_e$(S II).}

\figcaption[fig9.ps] {The abundance of (a) $^4$He, (b) D, (c) $^3$He 
and (d) $^7$Li
as a function of $\eta_{10}$ $\equiv$ 10$^{10}$ $\eta$, where $\eta$ is the 
baryon-to-photon 
number ratio, as given by the standard hot big bang nucleosynthesis model.
The theoretical dependences (solid lines) with 1$\sigma$ deviations
(dashed lines) are from Fiorentini et al. (1998). 
The abundances of D, $^3$He and $^7$Li are number ratios
relative to H. For $^4$He, the mass fraction $Y$ is shown. Our
value $Y_p$ = 0.2452$\pm$0.0015 gives $\eta$ = 
(4.7$^{+1.0}_{-0.8}$)$\times$10$^{-10}$
as shown by the solid vertical line. We show other data with 1$\sigma$ boxes.}

\figcaption[fig10.ps] {Joint fits to the baryon-to-photon number ratio, 
log $\eta_{10}$, and the equivalent number of light neutrino species $N_\nu$ 
using a $\chi^2$ analysis with the code developed by Fiorentini et al. (1998)
and Lisi et al. (1999) (a) for primordial abundance values $Y_p$ (this paper), 
(D/H)$_p$ (Burles \& Tytler 1998b) and ($^7$Li/H)$_p$ (Bonifacio \& Molaro 1997) 
and (b) for best choice of primordial abundance 
values $Y_p$ (this paper), (D/H)$_p$ (Levshakov et al. 1999) and ($^7$Li/H)$_p$ 
(Vauclair \& Charbonnel 1998). The minimum value $\chi^2_{min}$ = 0.001
in the latter case is achieved for log $\eta_{10}$ = 
0.648 ($\eta$ = 4.45$\times$10$^{-10}$) and $N_\nu$ = 3.006 (filled circle). 
Thin and thick solid lines are 1$\sigma$ and 2$\sigma$ deviations, respectively.
The experimental value $N_\nu$ = 2.993 (Caso et al. 1998) is shown by
dashed line.}

\clearpage

%
% Table 1 (line intensities in I Zw 18)
%
\begin{figure}
\plotfiddle{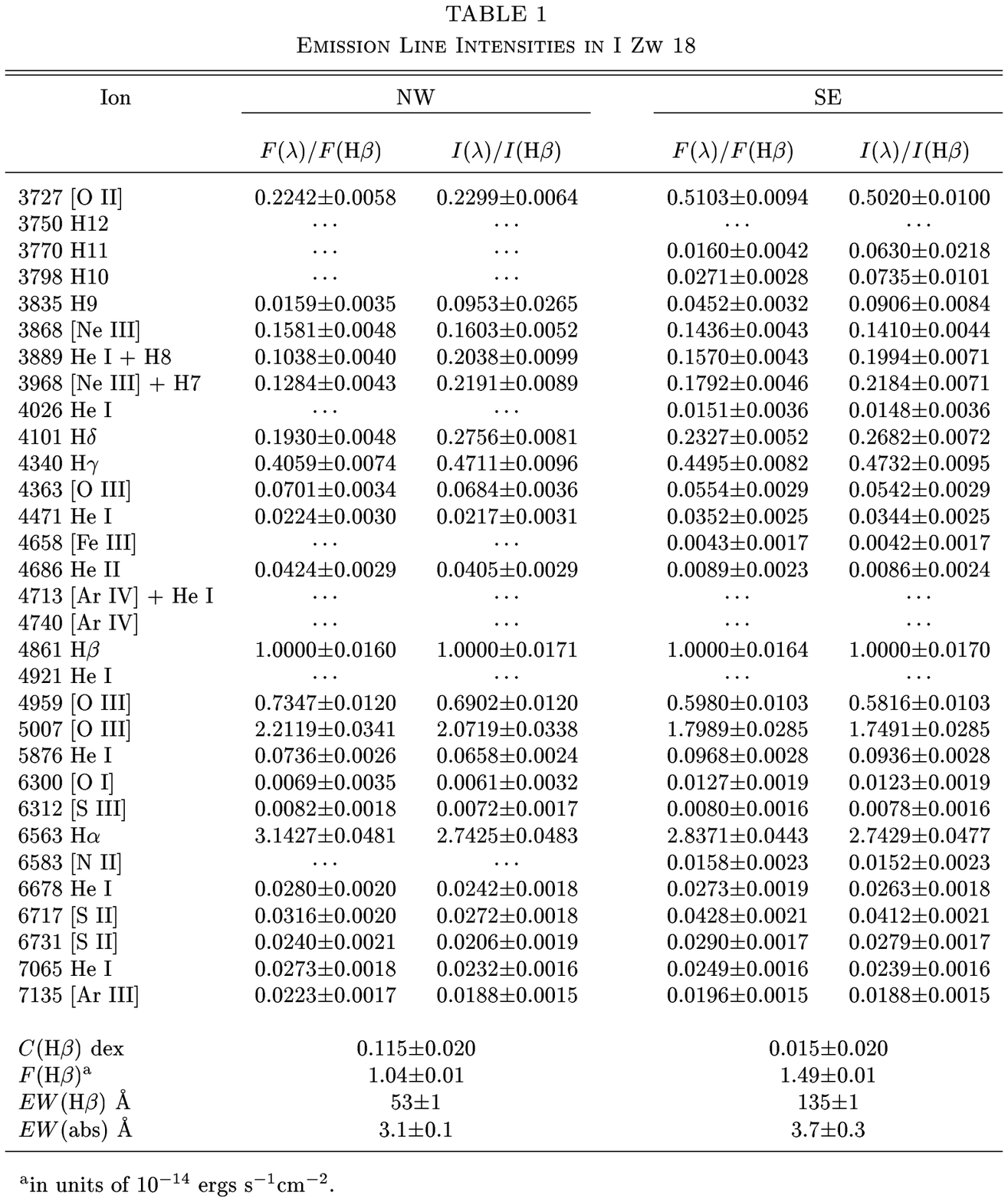}{0.cm}{0.}{90.}{90.}{-120.}{-350.}
\end{figure}

\clearpage

%
% Table 2a (line intensities in SBS 0335-052)
%
\begin{figure}
\plotfiddle{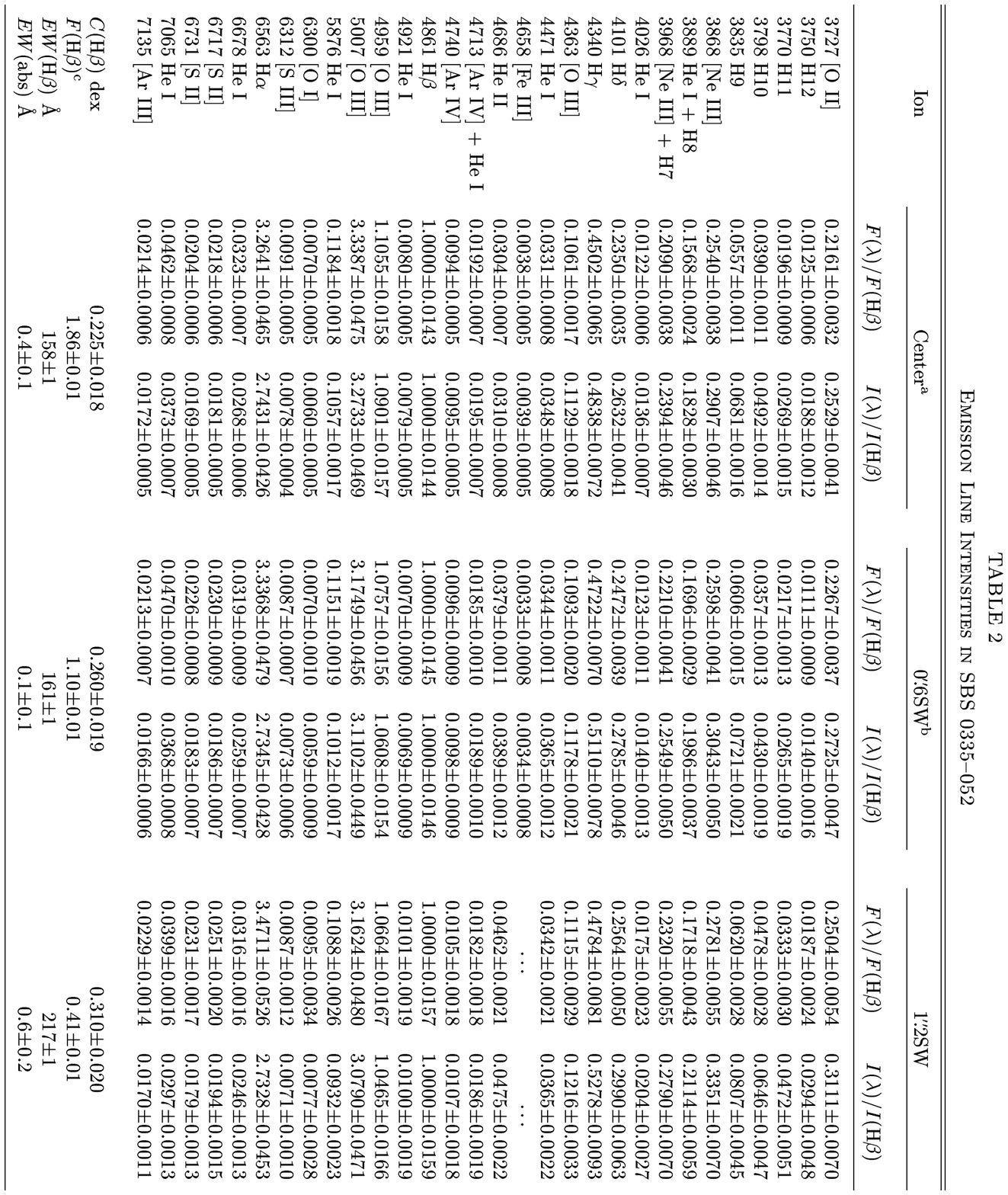}{0.cm}{180.}{90.}{90.}{440.}{450.}
\end{figure}

\clearpage

%
% Table 2b (line intensities in SBS 0335-052)
%
\begin{figure}
\plotfiddle{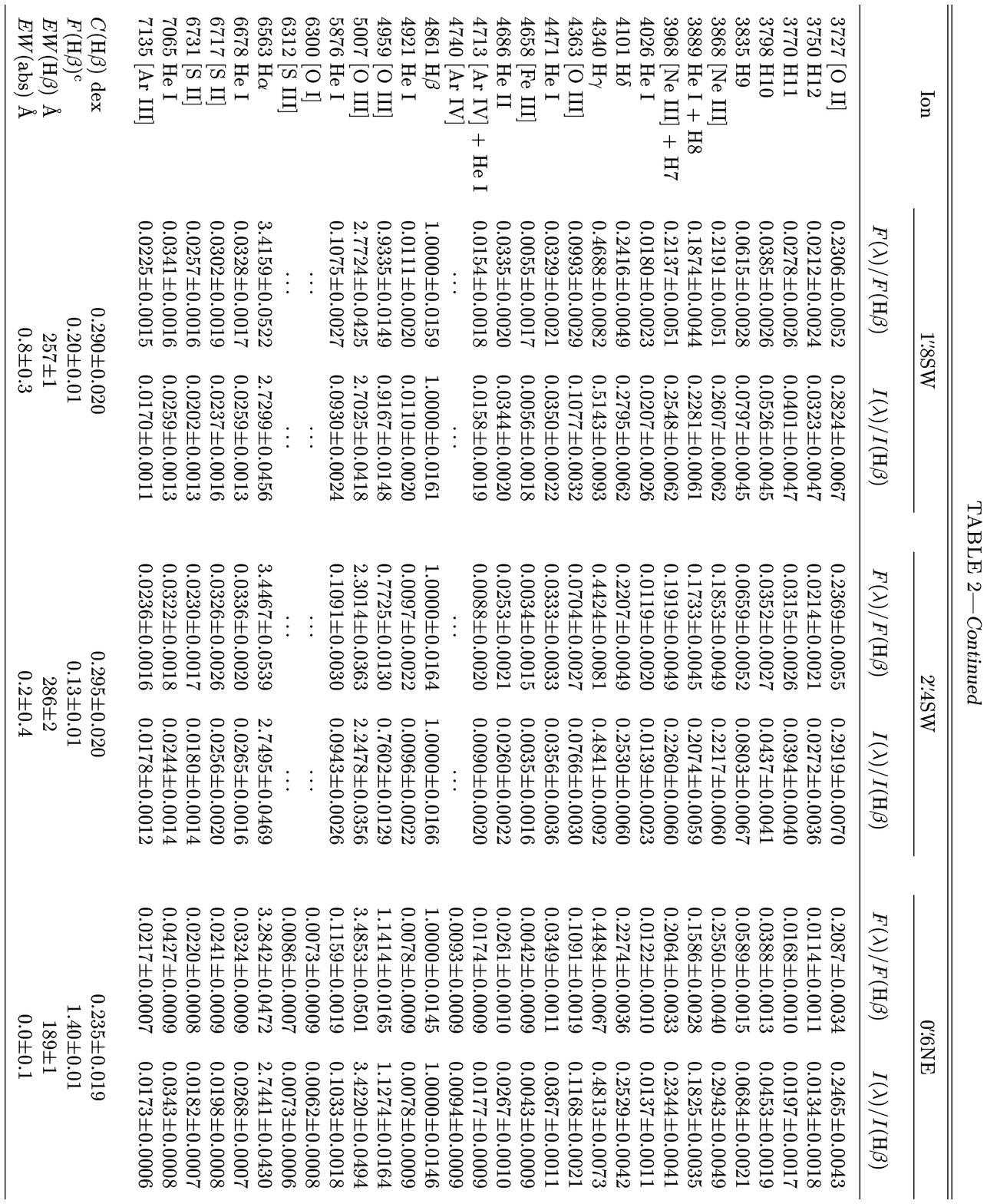}{0.cm}{180.}{90.}{90.}{440.}{450.}
\end{figure}

\clearpage

%
% Table 2c (line intensities in SBS 0335-052)
%
\begin{figure}
\plotfiddle{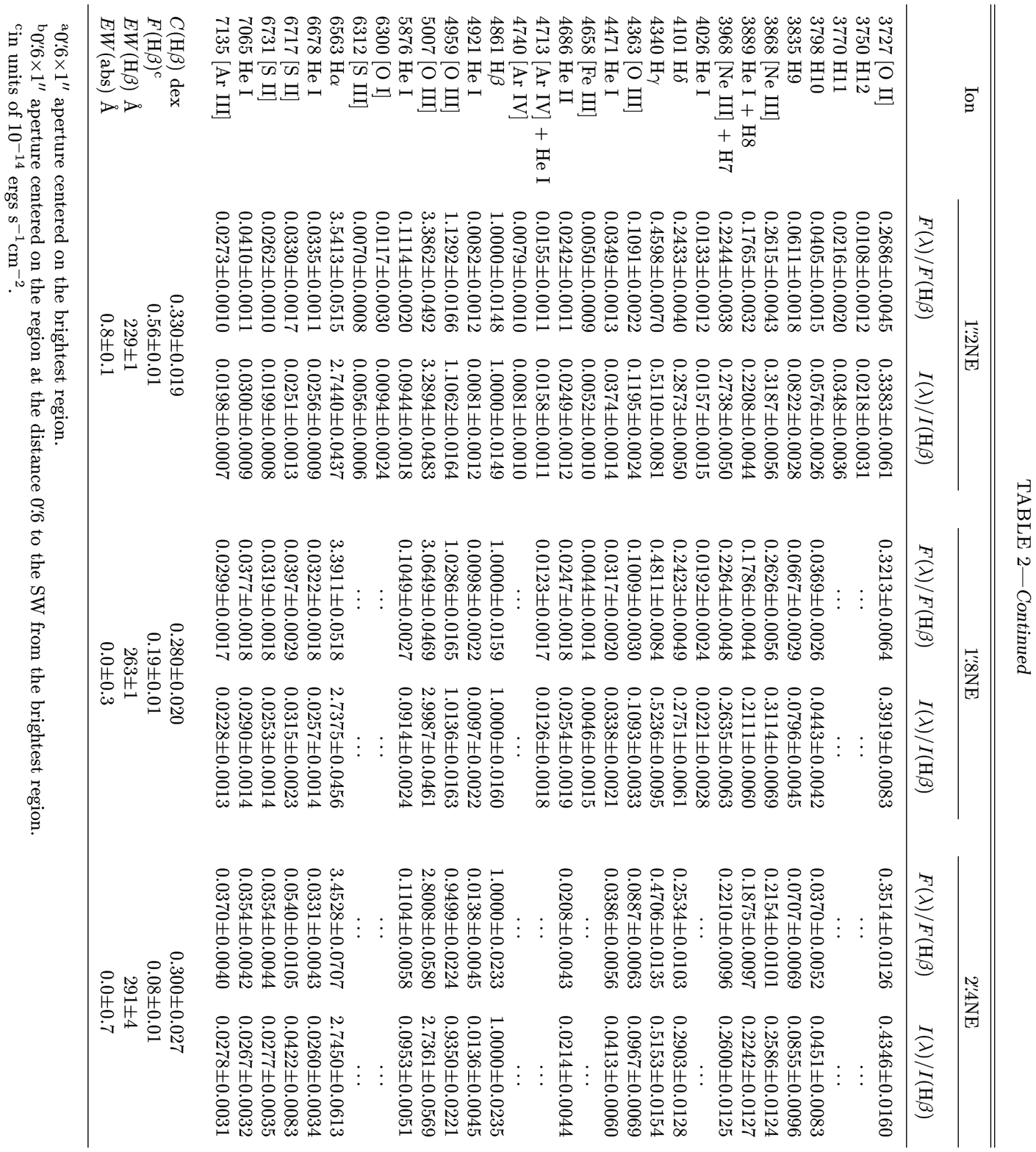}{0.cm}{180.}{90.}{90.}{440.}{450.}
\end{figure}

\clearpage
%*********************************
%   Table 3 Heavy element abundances in I Zw 18
%*********************************
\begin{figure}
\plotfiddle{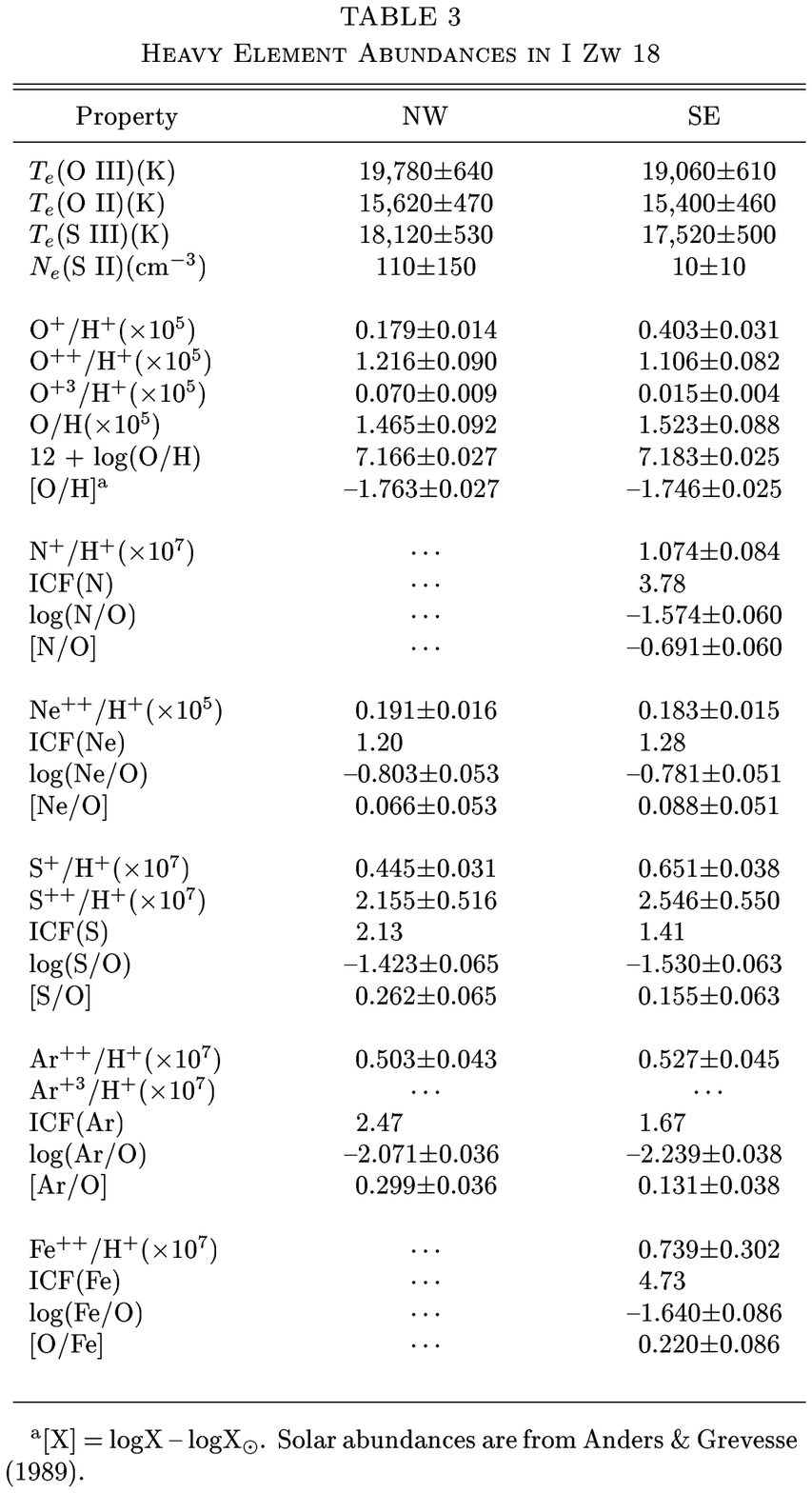}{0.cm}{0.}{90.}{90.}{-120.}{-350.}
\end{figure}

\clearpage
%*********************************
%   Table 4 Heavy element abundances in SBS 0335-052
%*********************************
\begin{figure}
\plotfiddle{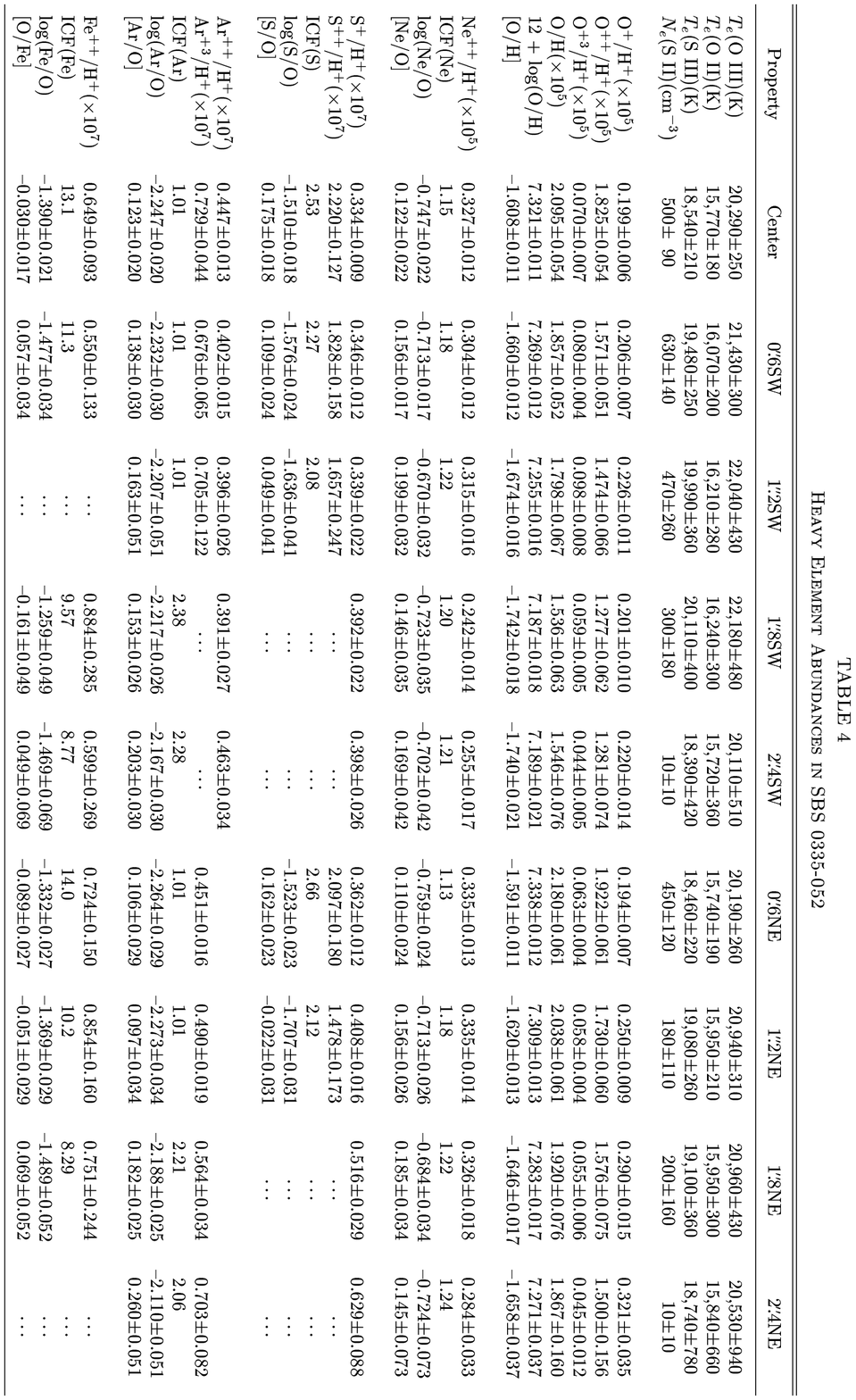}{0.cm}{180.}{90.}{90.}{400.}{420.}
\end{figure}

%*****************************************
%  Table 5 Helium abundance in I Zw 18
%*****************************************
\clearpage

\begin{figure}
\plotfiddle{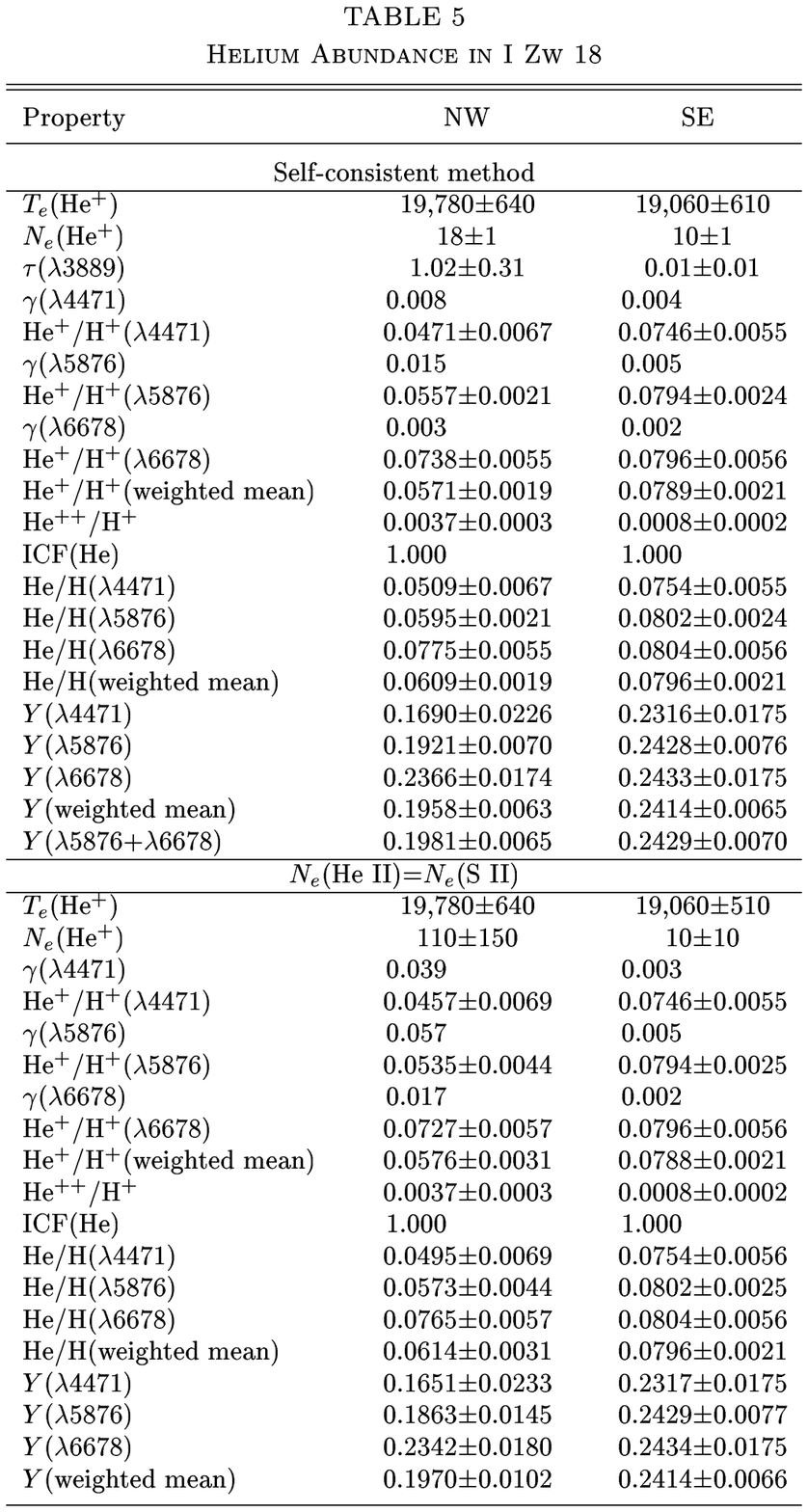}{0.cm}{0.}{90.}{90.}{-120.}{-350.}
\end{figure}

\clearpage

%*****************************************
% Table 6 Helium abundance in SBS 0335-052
%*****************************************
\begin{figure}
\plotfiddle{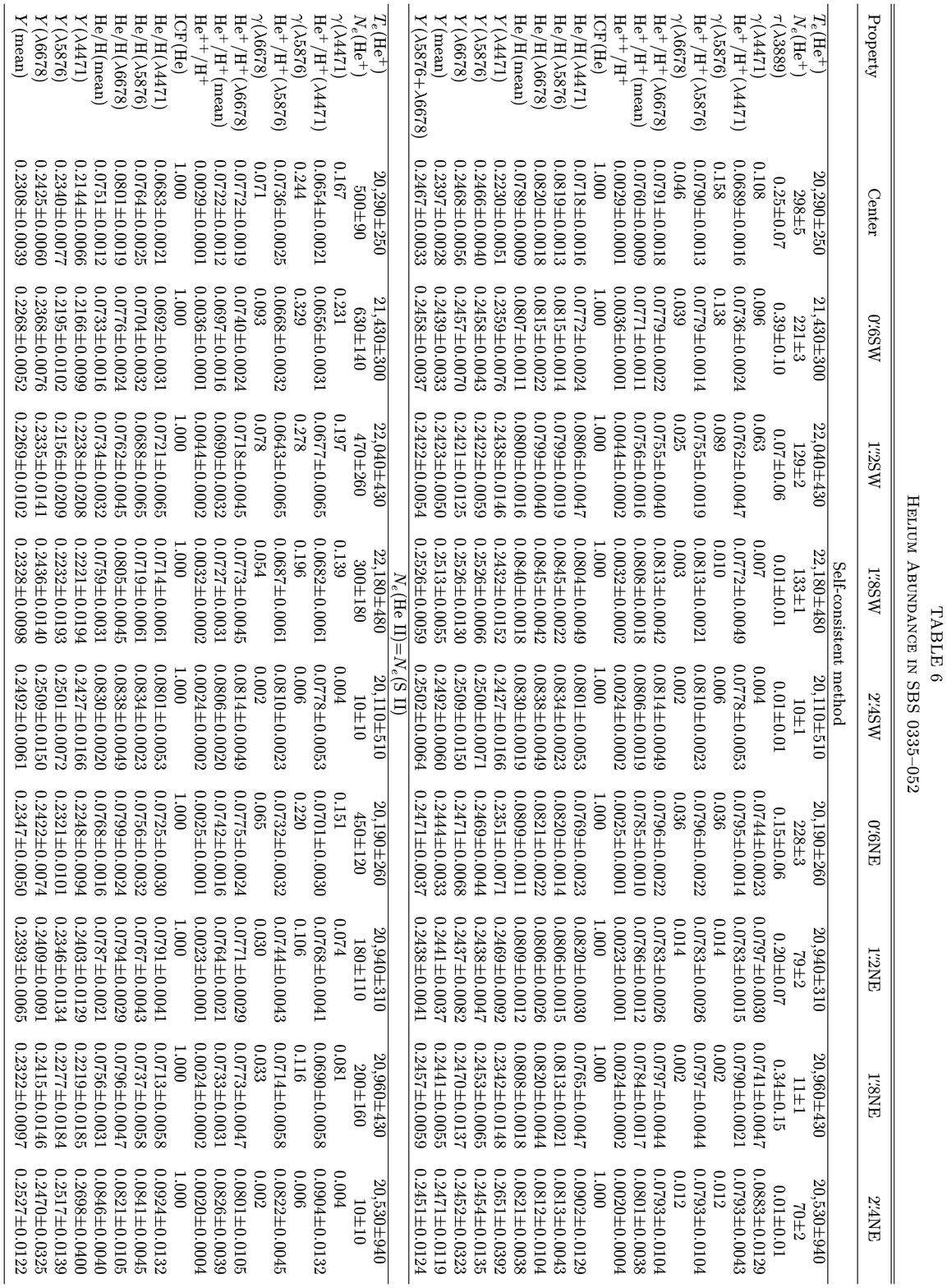}{0.cm}{180.}{80.}{80.}{380.}{380.}
\end{figure}

\clearpage
%******************************
%   Table 7 Weighted mean He abundances
%******************************
\begin{figure}
\plotfiddle{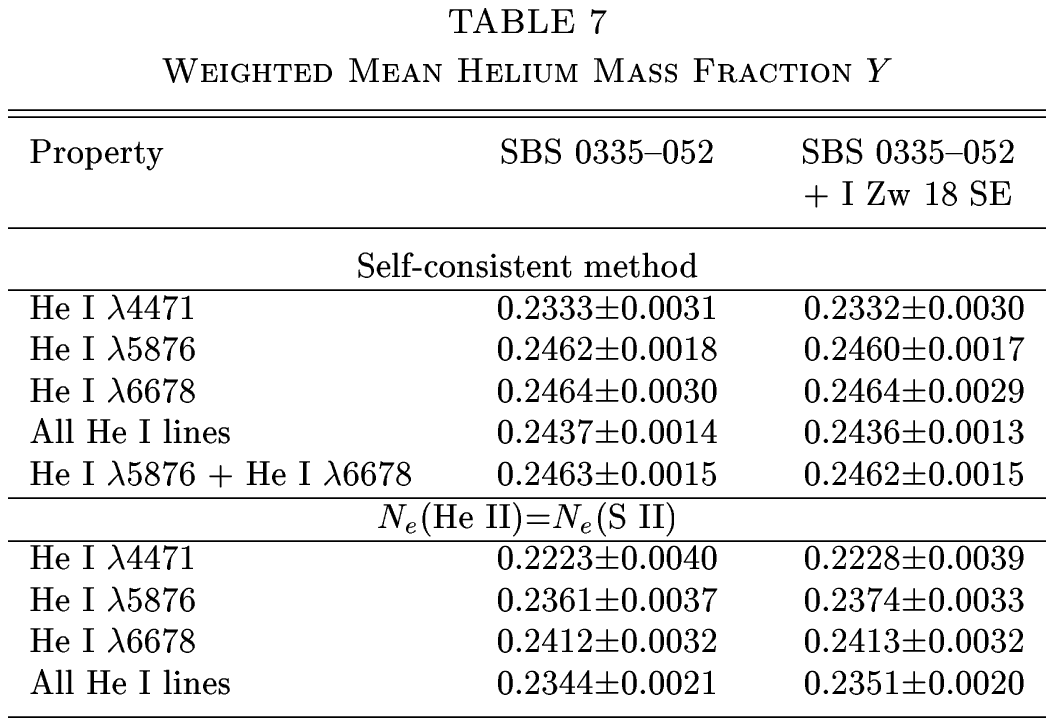}{0.cm}{0.}{100.}{100.}{-180.}{-400.}
\end{figure}


\clearpage

\begin{references}

\reference {} Aller, L. H. 1984, Physics of Thermal Gaseous Nebulae 
(Dordrecht: Reidel), p. 126
\reference {} Anders, E., \& Grevesse, N. 1989, Geochim. Cosmochim. Acta,
   53, 197
\reference {} Benjamin, R. A., Skillman, E. D., \& Smits, D. P. 1999,
\apj, in press; (astro-ph/9810087)
%\reference {} Bludman, S. A. 1998, \apj, 508, 535
\reference {} Bi, H., \& Davidsen, A. F. 1997, \apj, 479, 523
\reference {} Bohlin, R. C. 1996, \aj, 111, 1743
\reference {} Bonifacio, P. \& Molaro, P. 1997, \mnras, 285, 847
\reference {} Brocklehurst, M. 1972, \mnras, 157, 211
\reference {} Burles, S., Nollett, K. M., Truran, J. N., \& Turner, M. S.
1999, Phys. Rev. Lett, in press; (astro-ph/9901157)
\reference {} Burles, S., \& Tytler, D. 1998a, \apj, 499, 699
\reference {} ---------.\ 1998b, \apj, 507, 732
\reference {} Campbell, A., Terlevich, R., \& Melnick, J. 1986,
\mnras, 223, 811
\reference {} Caso, C. et al. (Particle Data Group) 1998, Eur. J. Phys., C3, 1
\reference {} Copi, C. J., Schramm, D. N., \& 
Turner, M. S. 1995, Science, 267, 192
\reference {} Esteban, C., Peimbert, M., Torres-Peimbert, S., \&
Escalante, V. 1998, \mnras, 295, 401
\reference {} Fiorentini, G., Lisi, E., Sarkar, S., Villante, F. L. 1998,
   Phys. Rev. D, 58, 063506
\reference {} Garnett, D. R. 1992, \aj, 103, 1330
\reference {} Izotov, Y. I., \& Thuan, T. X. 1998a, \apj, 500, 188
\reference {} ---------.\ 1998b, \apj, 497, 227
\reference {} ---------.\ 1999, \apj, 511, 639
\reference {}  Izotov, Y. I., Thuan, T. X., \& Lipovetsky, V. A. 1994, \apj, 
435, 647 (ITL94)
\reference {}  ---------.\ 1997, \apjs, 108, 1 (ITL97)
\reference {} Izotov, Y. I., Lipovetsky, V. A., Chaffee, F. H., 
Foltz, C. B., Guseva, N. G., \& Kniazev, A. Y. 1997a, \apj, 476, 698
\reference {} Izotov, Y. I., Foltz, C. B., Green, R. F., Guseva, N. G., \&
Thuan, T. X. 1997b, \apj, 487, L37
\reference {} Jaschek, M, Andrillat, Y., Houziaux, L., \& Jaschek, C. 1994,
\aap, 282, 911
\reference {} Kingdon, J., \& Ferland, G. J. 1995, \apj, 442, 714
\reference {}  Kunth, D., \& Sargent, W. L. W. 1983, \apj, 273, 81
\reference {} Legrand, F., Kunth, D., Roy, J.-R., Mas-Hesse, J. M.,
\& Walsh, J. R. 1997, \aap, 326, L17
\reference {} Leone, F., \& Lanzafame, A. C. 1998, \aap, 330, 306
\reference {} Levshakov, S. A., Kegel, W. H., \& Takahara, F. 1998,
\apj, 499, L1
\reference {} ---------.\ 1999, \mnras, 302, 707
%\reference {} Levshakov, S. A., Tytler, D., \& Burles, S. 1999, \aj, in press;
%(astro-ph/9812114)
\reference {} Lipovetsky, V. A., Chaffee, F. H., Izotov, Y. I.,
Foltz, C. B., Kniazev, A. Y., \& Hopp, U. 1999, \apj, July 1
\reference {} Lisi, E., Sarkar, S., \& Villante, F. L. 1999,
Phys. Rev. D, in press; (hep-ph/9901404)
%\reference {} Lopez, R. E., \& Turner, M. S. 1998, Phys. Rev. D, in press;
%(astro-ph/9807279)
\reference {} Maeder, A. 1992, \aap, 264, 105
\reference {} Martin, C. 1996, \apj, 465, 680
\reference {} Melnick, J., Heydari-Malayeri, M., \& Leisy, P. 1992, \aap,
253, 16
\reference {} Oke, J. B. 1990, \aj, 99, 1621
\reference {} Oke, J. B., Cohen, J. G., Carr, M., Cromer, J., Dingizian, A.,
Harris, F. H., Labrecque, S., Lucinio, R., Schall, W., Epps, H., \& Miller, J.
1995, \pasp, 107, 375
\reference {} Olive, K. A., \& Steigman, G. 1995, \apjs, 97, 49
\reference {} Olive, K. A., Skillman, E. D., \& Steigman, G. 1997,
\apj, 483, 788 (OSS97)
\reference {} Olofsson, K. 1995, \aaps, 111, 57
\reference {} Osterbrock, D. E., Tran, H. D., \& Veilleux, S. 1992,
    \apj, 389, 305
\reference {} Pagel, B. E. J., Simonson, E. A., Terlevich, R. J., \&
Edmunds, M. G. 1992, \mnras, 255, 325
\reference {} Pagel, B. E. J., Terlevich, R. J., \& Melnick, J. 
1986, \pasp, 98, 1005
\reference {} Papaderos, P., Izotov, Y. I., Fricke, K. J., Thuan, T. X.,
Guseva, N. G. 1998, \aap, 338, 43
\reference {} Peimbert, M. 1996, Rev. Mex. Astr. Astrof., 4, 55
\reference {} Peimbert, M., \& Torres-Peimbert, S. 1974, \apj, 193, 327
\reference {} ---------.\ 1976, \apj, 203, 581
\reference {} Pinsonneault, M. H., Walker, T. P., Steigman, G., \&
Naranyanan, V. K. 1998, \apj, in press; (astro-ph/9803073)
\reference {} Rauch, M., Miralda-Escud\'e, J., Sargent, W. L. W., 
Barlow, T. A., Weinberg, D. H., Hernquist, L., Katz, N., Cen, R., \&
Ostriker, J. P. 1997, \apj, 489, 7
\reference {} Robbins, R. R. 1968, \apj, 151, 511
\reference {} Rood, R. T., Bania, T. M., Balser, D. S., \& Wilson, T. L.
1998, Space Sci. Rev., 84, 185
\reference {} Sarkar, S. 1996, Rep. Prog. Phys., 59, 1493
\reference {} Skillman, E., \& Kennicutt, R. C., Jr. 1993, \apj, 411, 655 (SK93)
\reference {} Smits, D. P. 1996, \mnras, 278, 683
\reference {} Songaila, A., Wampler, E. J., \& Cowie, L. L. 1997, \nat,
    385, 137
\reference {} Stasi\'nska, G. 1990, \aaps, 83, 501
\reference {} Steigman, G., Schramm, D., \& Gunn, J. 1977, 
Phys. Lett. B, 66, 202
\reference {} Thuan, T. X., \& Izotov, Y. I. 1997, \apj, 489, 623
\reference {} Thuan, T. X., Izotov, Y. I., \& Lipovetsky, V. A. 1995, 
\apj, 445, 108
\reference {} ---------.\ 1996, \apj, 463, 120
\reference {} ---------.\ 1997, \apj, 477, 661
\reference {} Timmes, F. X., Woosley, S. E., \& Weaver, T. A. 1995,
\apjs, 98, 617
\reference {} Tytler, D., Burles, S., Lu, L., Fan, X.-M., \& Wolfe, A. 1999,
\aj, 117, 63
\reference {} van Zee, L., Westpfahl, D., Haynes, M. P., \& Salzer, J. J. 1998, 
\aj, 115, 1000
\reference {} Vauclair, S., \& Charbonnel, C. 1998, \apj, 502, 372
\reference {} V\'ilchez, J. M., \& Iglesias-P\'aramo, J. 1998, \apj, 508, 248
\reference {} V\'ilchez, J. M., \& Pagel, B. E. J. 1988, \mnras, 231, 257
\reference {} Walker, T. P., Steigman, G., Kang, H.-S., Schramm, D. M., \&
Olive, K. A. 1991, \apj, 376, 51
\reference {} Weinberg, D. H., Miralda-Escud\'e, J., Hernquist, L., \&
Katz, N. 1997, \apj, 490, 564
\reference {} Whitford, A. E. 1958, \aj, 63, 201
\reference {} Zhang, Y., Meiksin, A., Anninos, P., \& Norman, M. L. 1998,
\apj, 495, 63
\end{references}
\end{document}